\documentclass[aps,twocolumn,letter,showpacs,prepr intnumbers,amsmath,amssymb,superscriptaddress,bibnotes]{revtex4-1}

\usepackage{graphicx}
\usepackage{dcolumn}
\usepackage{bm}
\usepackage{textcomp}
\usepackage{amsmath}
\usepackage{amssymb}
\usepackage{color}
\usepackage{soul}
\usepackage{natbib}
\usepackage{dsfont}

\newcommand{\pals}{Pt$_{1-x}$Au$_{x}$LuSb}
\newcommand{\pbls}{Pt$_{7/8}$Au$_{1/8}$LuSb}

\newcommand{\pdls}{Pt$_{1/2}$Au$_{1/2}$LuSb}

\begin{document}
\title{Identifying the fingerprints of topological states by tuning magnetoresistance in a semimetal: the case of topological half-Heusler Pt$_{1-x}$Au$_{x}$LuSb }

\author{Shouvik Chatterjee}
\email[Authors to whom correspondence should be addressed: ]{shouvik.chatterjee@tifr.res.in, cjpalm@ucsb.edu}
\affiliation{Department of Electrical and Computer Engineering, University of California, Santa Barbara, CA 93106, USA}
\affiliation{Department of Condensed Matter Physics and Materials Science, Tata Institute of Fundamental Research, Homi Bhabha Road, Mumbai 400005, India}

\author{Felipe Crasto de Lima}
\affiliation{Department of Materials Science and Engineering, University of Delaware, Newark, DE 19716, USA}
\affiliation{Instituto de F\'isica, Universidade Federal de Uberl\^andia, C.P. 593, 38400-592, Uberl\^andia, MG, Brazil}

\author{John A. Logan}
\affiliation{Materials Department, University of California, Santa Barbara, CA 93106, USA}

\author{Yuan Fang}
\affiliation{Department of Physics and Astronomy, Stony Brook University, Stony Brook, New York, 11974, USA}

\author{Hadass Inbar}
\affiliation{Materials Department, University of California, Santa Barbara, CA 93106, USA}

\author{Aranya Goswami}
\affiliation{Department of Electrical and Computer Engineering, University of California, Santa Barbara, CA 93106, USA}

\author{Connor Dempsey}
\affiliation{Department of Electrical and Computer Engineering, University of California, Santa Barbara, CA 93106, USA}

\author{Jason Dong}
\affiliation{Materials Department, University of California, Santa Barbara, CA 93106, USA}

\author{Shoaib Khalid}
\affiliation{Department of Physics and Astronomy, University of Delaware, Newark, DE 19716, USA}
\affiliation{Department of Materials Science and Engineering, University of Delaware, Newark, DE 19716, USA}

\author{Tobias Brown-Heft}
\affiliation{Materials Department, University of California, Santa Barbara, CA 93106, USA}

\author{Yu-Hao Chang}
\affiliation{Materials Department, University of California, Santa Barbara, CA 93106, USA}

\author{Taozhi Guo}
\affiliation{Department of Physics, University of California, Santa Barbara, CA 93106, USA}

\author{Daniel Pennachio}
\affiliation{Materials Department, University of California, Santa Barbara, CA 93106, USA}

\author{Nathaniel Wilson}
\affiliation{Materials Department, University of California, Santa Barbara, CA 93106, USA}

\author{Shalinee Chikara}
\affiliation{National High Magnetic Field Laboratory, Tallahassee, FL 32310, USA}

\author{Alexey Suslov}
\affiliation{National High Magnetic Field Laboratory, Tallahassee, FL 32310, USA}

\author{Alexei V. Fedorov}
\affiliation{Advanced Light Source, Lawrence Berkeley National Laboratory, Berkeley, CA 94720, USA}

\author{Dan Read}
\affiliation{Department of Electrical and Computer Engineering, University of California, Santa Barbara, CA 93106, USA}
\affiliation{School of Physics and Astronomy, Cardiff University, Cardiff CF24 3AA, UK}

\author{Jennifer Cano}
\affiliation{Department of Physics and Astronomy, Stony Brook University, Stony Brook, New York, 11974, USA}
\affiliation{Center for Computational Quantum Physics, Flatiron Institute, New York, New York, 10010, USA}

\author{Anderson Janotti}
\affiliation{Department of Materials Science and Engineering, University of Delaware, Newark, DE 19716, USA}

\author{Christopher J. Palmstr\o m}
\email[Authors to whom correspondence should be addressed: ]{shouvik.chatterjee@tifr.res.in, cjpalm@ucsb.edu}
\affiliation{Department of Electrical and Computer Engineering, University of California, Santa Barbara, CA 93106, USA}
\affiliation{Materials Department, University of California, Santa Barbara, CA 93106, USA}
\affiliation{California NanoSystems Institute, University of California, Santa Barbara, California 93106, USA}

\begin{abstract}

Topological materials often exhibit remarkably linear, non-saturating magnetoresistance (LMR), which is both of scientific and technological importance. However, the role of topologically non-trivial states in the emergence of such a behaviour has eluded clear demonstration in experiments. Here, by reducing the coupling between the topological surface states (TSS) and the bulk carriers we controllably tune the LMR behavior in \pals\/ into distinct plateaus in Hall resistance, which we show arise from a quantum Hall phase. This allowed us to reveal how smearing of the Landau levels, which otherwise give rise to a quantum Hall phase, results in an LMR behavior due to strong interaction between the TSS with a positive $g$-factor and the bulk carriers. We establish that controlling the coupling strength between the surface and the bulk carriers in topological materials can bring about dramatic changes in their magnetotransport behavior.  In addition, our work outlines a strategy to reveal macroscopic physical observables of TSS in compounds with a semi-metallic bulk band structure, as is the case in multi-functional Heusler compounds, thereby opening up opportunities for their utilization in hybrid quantum structures.

\end{abstract}
\maketitle

\section{Introduction}
In conventional materials, resistance under the application of magnetic field is expected to have a quadratic dependence on the magnetic field strength ($H$), saturating at high field values ($\nu\mu_{0} H \geq 1$, $\nu$ - carrier mobility, $\mu_{0}$ - vacuum permeability), except for systems where electron and hole densities are perfectly compensated \cite{pippard1989magnetoresistance}. In contrast, magnetoresistance in a number of topological materials exhibits remarkably linear dependence on the magnetic field strength \cite{leahy2018nonsaturating, wang2012room, yan2013large, zhang2012magneto, narayanan2015linear}. Several theoretical proposals have been put forward to explain such a phenomenon, which can be classified into two distinct categories. One set of theoretical proposals invokes generic semi-classical considerations such as spatial mobility fluctuations due to nanoscale inhomogeneity \cite{xu1997large, parish2003non}, guiding center diffusion in a weak disorder \cite{song2015linear}, and long-range correlated disorder \cite{nandi2018signatures}. In contrast, the other set of proposals rely on some characteristics of the topological states such as the extreme quantum limit of linearly dispersive states where only the lowest Landau level is occupied \cite{abrikosov1998quantum}, Zeeman splitting of the Landau levels arising from the topological surface states (TSS) \cite{wang2012linear}, and quantum coherence effects in the electronic states with strong spin-orbit coupling \cite{zhang2012magneto, assaf2013linear}. However, the mechanism behind the origin of such a phenomenon in topological materials and the role of TSS have remained unclear due to the difficulty in distinguishing between these disparate scenarios in experiments. Furthermore, although a  plethora of interesting topological states have been predicted in semi-metallic systems \cite{burkov2016topological,armitage2018weyl,klemenz2019topological}, including multi-functional Heusler compounds\cite{chadov2010tunable,lin2010half,yang2017prediction,wang2016time,chang2017topological,chang2016room}, the inability to disentangle the contribution of the topological states from the bulk carriers imposes serious constraints on their utilization in functional devices. To address these outstanding issues, we have adopted a strategy of controllably tuning the linear magnetoresistance behavior in semi-metallic, half-Heusler epitaxial thin films into a quantum Hall-like phase that arises from the TSS, which is achieved by modifying the coupling between the surface and the bulk carriers.

\begin{figure}
\centering
\includegraphics[width=0.9\linewidth]{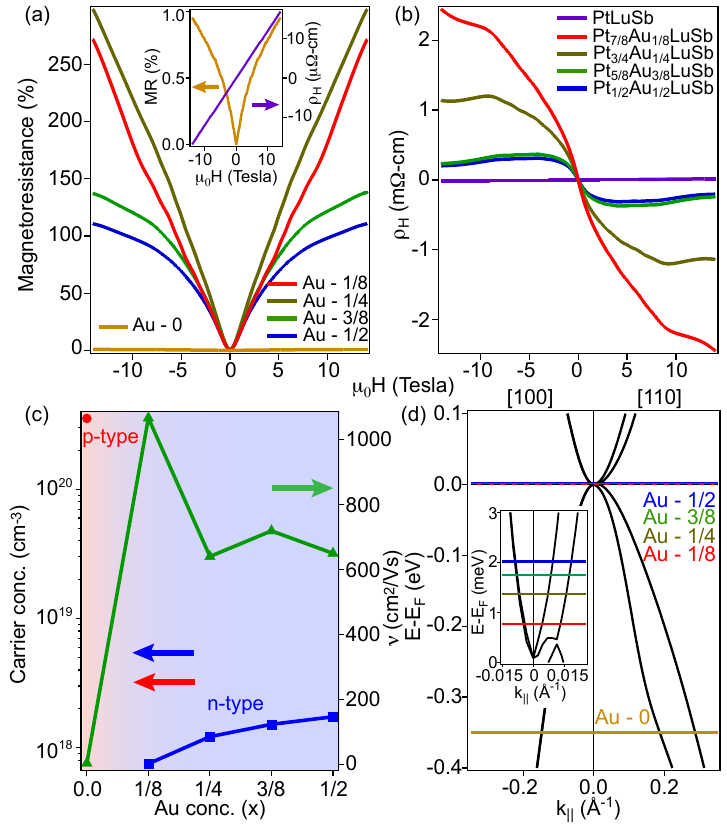}
\caption{Transport properties of Pt$_{1-x}$Au$_{x}$LuSb thin films. (a) Magnetoresistance and (b) Hall resistance at 2K as a function of Au concentration in \pals\/ thin films. Inset in (a) shows the magnetoresistance and Hall resistance of PtLuSb. (c) Carrier concentration and mobility ($\nu$) as a function of Au concentration in \pals\/, estimated from the Hall coefficient. (d) Position of the Fermi level as a function of Au concentration (see text).  }
\label{fig:Transport}
\end{figure}

 In the half-Heusler compound PtLuSb, a band inversion is predicted between the $\Gamma_{8}$ (p-type character) and the $\Gamma_{6}$ (s-type character) manifolds resulting in its topologically non-trivial character \cite{chadov2010tunable,lin2010half,logan2016observation}. In such cases, bulk-boundary correspondence guarantees the existence of TSS that have been observed in recent angle-resolved photoemission spectroscopy (ARPES) measurements \cite{logan2016observation}. However, the presence of a large number of bulk carriers, due to the semimetallic nature of these compounds, have precluded the identification of macroscopic physical observables of the TSS in any Heusler compound till date. Intrinsic doping in these compounds, ascribed to the presence of anti-site defects \cite{yonggang2017natural}, further exacerbates the problem. In PtLuSb, this results in the Fermi level to lie deep inside the valence band in contradiction to the density functional theory (DFT) calculations. From the DFT calculations the chemical potential is expected to lie at the quadratic band touching point ($\Gamma_{8}$), at the top of the bulk valence band (Fig.~\ref{fig:Transport}(d)) \cite{logan2016observation, patel2014surface}. By substituting a few Pt atoms with Au in \pals\/ thin films, where Au atoms are expected to contribute one extra electron compared to Pt, we are able to lower the bulk carrier concentration by more than two orders of magnitude compared to the parent compound. The chemical potential in the bulk in such cases is expected to lie close to the bottom of the conduction band, as shown in Fig.~\ref{fig:Transport}(d).

In \pals\/ samples with low Au concentration, we show how TSS interact strongly with the semi-localized charge carriers in the bulk leading to LMR. With an increase in the Au concentration the coupling between the surface states and the bulk carriers is reduced. Consequently, the LMR behavior is transformed to reveal well-defined resistance plateaus as a function of magnetic field, which we argue arise from a quantum Hall phase. This has allowed us to identify the mechanism behind the origin of LMR in topological \pals\/ thin films and elucidate the role of TSS and their coupling with the localized bulk carriers. Furthermore, our magneto-transport results on substitution alloyed thin films establishes a strategy to reveal properties of TSS in semi-metallic systems. In particular, for the Heusler compounds this can potentially open up the possibility of fabricating high-quality hybrid devices \cite{mellnik2014spin, qi2011topological} by combining the topological properties with magnetism \cite{de1983new} and superconductivity \cite{klimczuk2012superconductivity, nakajima2015topological} within the same material system, utilizing their multi-functional, yet structurally similar nature \cite{hinterleitner2019thermoelectric,wollmann2017heusler, graf2011simple, palmstrom2016heusler}.

\section{Results}
\subsection{Thin film growth and magnetotransport at low fields}
Epitaxial thin films of 15 nm thick \pals\/ were synthesized on InSb buffer layer on GaAs(001) substrates. Details of the growth process and film characterization can be found in the Methods section and in the Supplementary Material \cite{suppl}. The addition of Au is found to have a dramatic effect on the charge transport in \pals\/ thin films. Magnetoresistance ($\Delta \rho_{xx} (B) / \rho_{xx} (0)$) in substitution alloyed thin films increases by more than two orders of magnitude compared to PtLuSb, as shown in Fig.~\ref{fig:Transport}(a). Magnetoresistance in thin films with a low Au concentration (x =1/8) shows a remarkably linear non-saturating magnetoresistance behavior. When the gold concentration is increased (x=3/8, 1/2), it evolves into a sub-linear magnetic field dependence of the magnetoresistance  that begins to saturate at high magnetic fields. The carrier concentration, as estimated from the low-field Hall measurements, drops from 3.53$\times$10$^{20}$ cm$^{-3}$ in PtLuSb to 7.53$\times$10$^{17}$ cm$^{-3}$ in Pt$_{7/8}$Au$_{1/8}$LuSb, a change by almost three orders of magnitude on the addition of gold. Low-field Hall measurements also indicate a change in the carrier type from predominantly p-type to n-type in the gold alloyed samples. The bulk band structure obtained from the density functional theory (DFT) calculations coupled with the estimated carrier concentration (details can be found in the Supplementary Material \cite{suppl}) indicate that the Fermi level lies very close to the four-fold degenerate $\Gamma_{8}$ point in the gold doped samples, as shown in Fig.~\ref{fig:Transport}(d).

\subsection{Electronic structure measured by ARPES}
\begin{figure}[htp]
\centering
\includegraphics[width=0.9\linewidth]{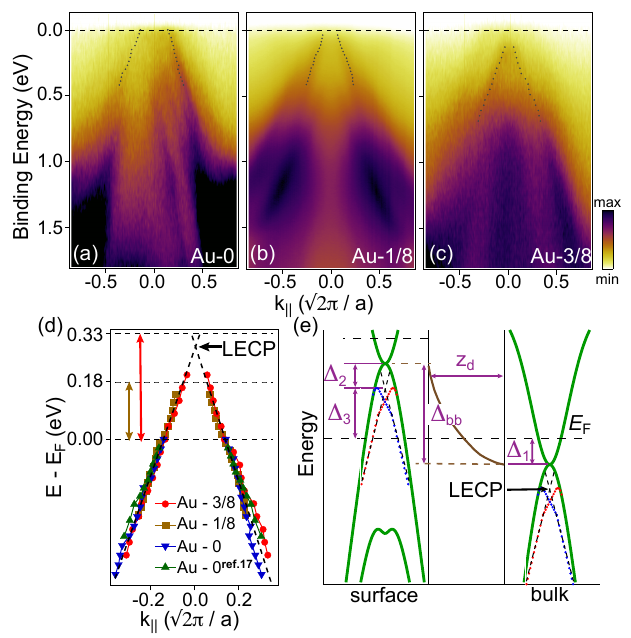}
\caption{Topological surface state in Pt$_{1-x}$Au$_{x}$LuSb thin films. ARPES measurements along $\overline{\mathit{\Gamma}}$ - $\overline{X}$ in \pals\/ thin films close to the bulk $\Gamma$ point for (a) x= 0 (b) x= 1/8 (c) x= 3/8. (d) Extracted dispersion of the topological surface state (TSS) showing the shift in the Fermi level (E$_{F}$) with the addition of gold (Au). (e) Schematic of the upward band-bending observed in \pals\/ thin films. $\Delta_{1}$, $\Delta_{2}$, and $\Delta_{3}$ represent the position of the Fermi level with respect to $\Gamma_{8}$ in the bulk ($\Delta_{1}$), the position of the linear extrapolated crossing point (LECP) with respect to $\Gamma_{8}$ ($\Delta_{2}$), and the position of the LECP with respect to the Fermi level at the surface ($\Delta_{3}$), which are estimated from the Hall measurements along with the bulk DFT calculations, DFT slab calculations, and ARPES measurements, respectively. Band bending ($\Delta_{bb}$) is obtained as $\Delta_{bb} = \Delta_{1} + \Delta_{2} + \Delta_{3}$. Details about the estimation of band bending can be found in the Supplementary Material \cite{suppl}.}
\label{fig:ARPES}
\end{figure}

In addition to donating one extra electron per formula unit, the addition of gold is expected to maintain the band inversion similar to what has been predicted for the parent compound PtLuSb \cite{suppl}, thus providing a pathway to tune the occupation of the TSS. In our experiment, the changes in the electronic structure due to substitution alloying of gold in \pals\/ is directly observed in the angle-resolved photoemission spectroscopy (ARPES) measurements. In Fig.~\ref{fig:ARPES}, we show the ARPES data taken along $\overline{\mathit{\Gamma}}$ - $\overline{X}$ of the surface Brillouin zone (see Figs.~S6,S7 in \cite{suppl}) close to the bulk $\Gamma$ point for three different gold concentrations in \pals\/ with x = 0, 1/8, and 3/8. For the parent compound PtLuSb, the extracted dispersion of the TSS, shown in Fig.~\ref{fig:ARPES}(d), is identical to what has been observed in our previous work, where spin-momentum locking of the surface state was revealed by spin-resolved ARPES \cite{logan2016observation}. This confirms that we are measuring the same linearly dispersive TSS in our current set of samples. The two-dimensional nature of the TSS is established by the photon energy dependent measurements, shown in Fig.~S7 in \cite{suppl}. Upon the addition of gold, the chemical potential is shifted to higher energies but the dispersion of the TSS remains mostly unchanged (Fig.~\ref{fig:ARPES}). The measured Fermi velocity of the TSS is $v_{F}$ = 4.85 $\times$ 10$^{5}$ m/s, which is in close agreement with the predicted $v_{F}$ = 4.9 $\pm$ 0.3 $\times$ 10$^{5}$ m/s in the DFT slab calculations (see Fig.~S6 in \cite{suppl}). From the extracted dispersion of the TSS, we estimate the shift in the surface Fermi energy due to the addition of Au to be 180 meV and 330 meV in Pt$_{7/8}$Au$_{1/8}$LuSb and Pt$_{5/8}$Au$_{3/8}$LuSb, respectively, shown in Fig.~\ref{fig:ARPES}(d). 

\subsection{Massive Dirac fermions and surface band bending}
In Pt$_{5/8}$Au$_{3/8}$LuSb, the surface Fermi level is sufficiently raised to reveal the band gap in TSS (Fig.~\ref{fig:ARPES}(c)) indicative of the formation of massive Dirac fermions due to coupling between the TSS at opposite surfaces in \pals\/ thin films (see also Fig.~S7 in \cite{suppl}). The wavefunctions of TSS at opposite surfaces are expected to overlap and open a bandgap in thin films for thicknesses less than the critical thickness $\xi_{c}$, given by $\xi_{c}$ = $2\hbar v_{F}/\Delta$. Here, $\Delta$ is the charge excitation gap and $\xi$ is the surface penetration depth \cite{linder2009anomalous}. We estimate a bulk band gap of 18.72 meV in our 15 nm thick films due to quantum confinement from the \emph{k.p} fitting of the bulk band structure calculated from DFT (see Section IV, Fig.~S10 in \cite{suppl}). This results in a critical thickness $\xi_{c}$ $\approx$ 34 nm, which is larger than the thickness of our thin films (15 nm) indicating a significant overlap between the TSS wavefunctions at the opposite surfaces, in agreement with our ARPES data. A comparison between the Fermi levels obtained from the ARPES (surface) and Hall measurements (bulk) indicates an upward band-bending near the sample surface, as shown schematically in Fig.~\ref{fig:ARPES}(e), typically observed in semiconductors where the Fermi level is pinned at the surface \cite{brillson2010surfaces}. While the band-bending is negligible (3 meV) in PtLuSb, a steep band-bending of 225 meV and 123 meV is observed for Pt$_{7/8}$Au$_{1/8}$LuSb and Pt$_{5/8}$Au$_{3/8}$LuSb, respectively. Details about the estimation of band-bending can be found in the Supplementary Material \cite{suppl} (Section II, Table S1). The observed trend is likely due to the presence of surface/interfacial states coupled with low bulk carrier concentration in the gold doped samples. In contrast, in PtLuSb where the bulk carrier concentration is large, the band bending effect is indeed negligible. The presence of bulk bands crossing the Fermi level  at the surface in addition to the TSS due to the semi-metallic nature of \pals\/ is expected to result in a short depletion width. A depletion width shorter than the film thickness (15 nm) would result in different Fermi level estimations from the ARPES measurements that primarily probe the surface region and the bulk transport measurements, as is observed in \pals\/ thin films.
\begin{figure*}
\centering
\includegraphics[width=0.9\textwidth]{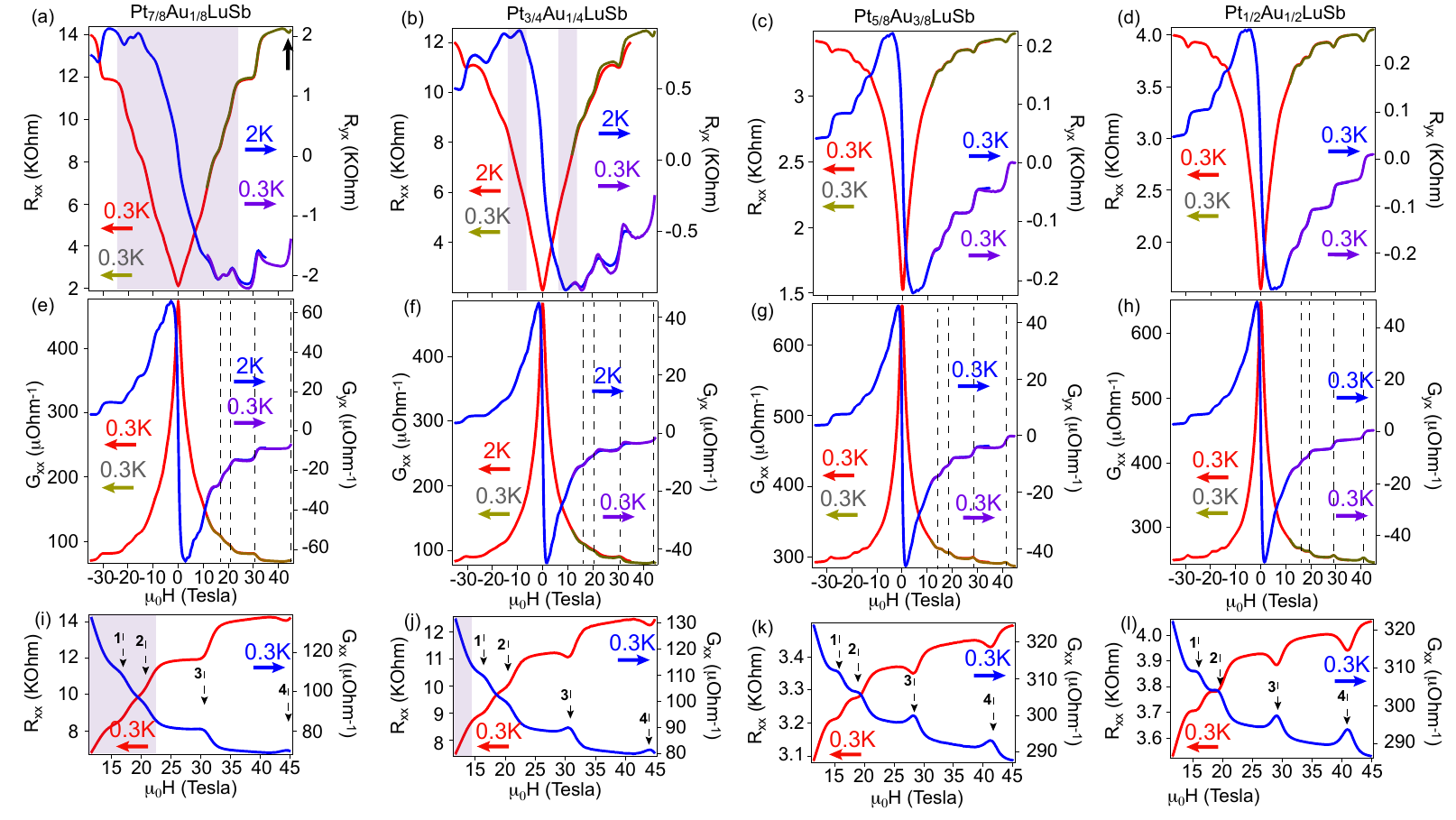}
\caption{Linear magnetoresistance behavior to a quantum Hall phase in a topological semi-metallic thin film. Longitudinal and Hall resistance in (a) Pt$_{7/8}$Au$_{1/8}$LuSb, (b) Pt$_{3/4}$Au$_{1/4}$LuSb, (c) Pt$_{5/8}$Au$_{3/8}$LuSb, (d) Pt$_{1/2}$Au$_{1/2}$LuSb. Longitudinal and Hall conductance in (e) Pt$_{7/8}$Au$_{1/8}$LuSb, (f) Pt$_{3/4}$Au$_{1/4}$LuSb, (g) Pt$_{5/8}$Au$_{3/8}$LuSb, (h) Pt$_{1/2}$Au$_{1/2}$LuSb. Correspondence between the local minima in longitudinal resistance and the local maxima in longitudinal conductance in (i) Pt$_{7/8}$Au$_{1/8}$LuSb, (j) Pt$_{3/4}$Au$_{1/4}$LuSb, (k) Pt$_{5/8}$Au$_{3/8}$LuSb, (l) Pt$_{1/2}$Au$_{1/2}$LuSb. Linear magnetoresistance behavior in Pt$_{7/8}$Au$_{1/8}$LuSb and in Pt$_{3/4}$Au$_{1/4}$LuSb is highlighted by a light violet background in (a),(i) and (b),(j), respectively. }
\label{fig:QH}
\end{figure*}
\subsection{Quantum Hall effect in \pals\/ thin films}
Having gained an understanding of both the bulk and the surface electronic structure in \pals\/ thin films, we turn towards their magnetotransport data, where the measurements were taken upto a high magnetic field value of 45 T. Remarkably linear non-saturating magnetoresistance is observed in Pt$_{7/8}$Au$_{1/8}$LuSb from very low magnetic field values up to a magnetic field of $\approx$22 T, above which it begins to deviate (Fig.~\ref{fig:QH}(a)). However, in Pt$_{3/4}$Au$_{1/4}$LuSb the LMR behavior is observed only between $\approx$6 and 14.5 T, where a crossover from a sub-linear to linear magnetoresistance happens at $\approx$ 6 T, shown in Fig.~\ref{fig:QH}(b). Such a crossover has been observed in a number of material systems with TSS \cite{wang2012room, singh2017linear} and is attributed to a reduction in surface-bulk coupling. In Pt$_{5/8}$Au$_{3/8}$LuSb and Pt$_{1/2}$Au$_{1/2}$LuSb, where the surface-bulk coupling is even weaker as discussed in the next section, linear magnetoresistance is completely absent, and the resistance exhibits a sub-linear magnetic field dependence throughout the measured field range (Fig.~\ref{fig:QH}(c)-(d)). 

Although the magnetoresistance behavior of \pals\/ films with different Au concentration appears very different, that they share a common underlying origin is established from their respective magnetoconductance plots $G_{xx/xy} = \frac{R_{xx/xy}}{R_{xx}^{2} +R_{xy}^{2}}$ , shown in Fig.~\ref{fig:QH}(e)-(h).  For all Au concentrations measured in this study, plateaus in $G_{xy}$ emerge at high magnetic fields above $\approx$10T, when the bulk quantum limit is reached (Fig.~\ref{fig:QH}(e)-(h), see also Section IV, Fig.~S11 in \cite{suppl}). The plateaus in $G_{xy}$ is accompanied by a corresponding local maxima in $G_{xx}$ as expected for quantum Hall states. However, in contrast to the traditional quantum Hall systems we observe local minima in magnetoresistance instead of local maxima. This could be explained by noting that in \pals\/, R$_{xy}$ $\ll$ R$_{xx}$, leading to an additional negative sign when converting to resistance using the conductance tensor \cite{ando2013topological}. The local minima in magnetoresistance indeed correspond to the local maxima in magnetoconductance, as shown in Fig.~\ref{fig:QH}(i)-(l). 

Moreover, the observed magnitude of R$_{xy}$/G$_{xy}$ is not quantized in units of (h/e$^{2}$)/(e$^{2}$/h). This is to be expected due to the semi-metallic nature of \pals\/ and also due to parallel conduction from InSb buffer layers in \pals\/ thin film structures, evident from positive magnetoresistance and non-linear Hall resistance (see Fig.~\ref{fig:Transport} (a)-(b)). This can be contrasted with the situation in near-surface InSb quantum wells where the \emph{n}-type quantum Hall effect is expected to be much better behaved with much less parallel conduction compared to \pals\/, yet exhibits R$_{xy}$ not being quantized at the expected values \cite{lei2019quantum}. In \pals\/, the situation is exacerbated due to the presence of multiple carriers of opposite polarity (n-type bulk carriers and p-type carriers from TSS) that leads to a change in sign of the Hall voltage with applied magnetic field (Fig.~\ref{fig:QH}). Therefore, although perfect quantization could not be achieved in \pals\/ thin film structures, the observation of well-defined plateaus in $G_{xy}$ and corresponding maxima in $G_{xx}$ is taken as evidence for the emergence of quantum Hall phase, which is ascribed to the presence of two-dimensional TSS. We will further discuss the origin of such a phase and rule out other possible alternate scenarios in Section III.

%
\begin{figure*}
\centering
\includegraphics[width=0.9\textwidth]{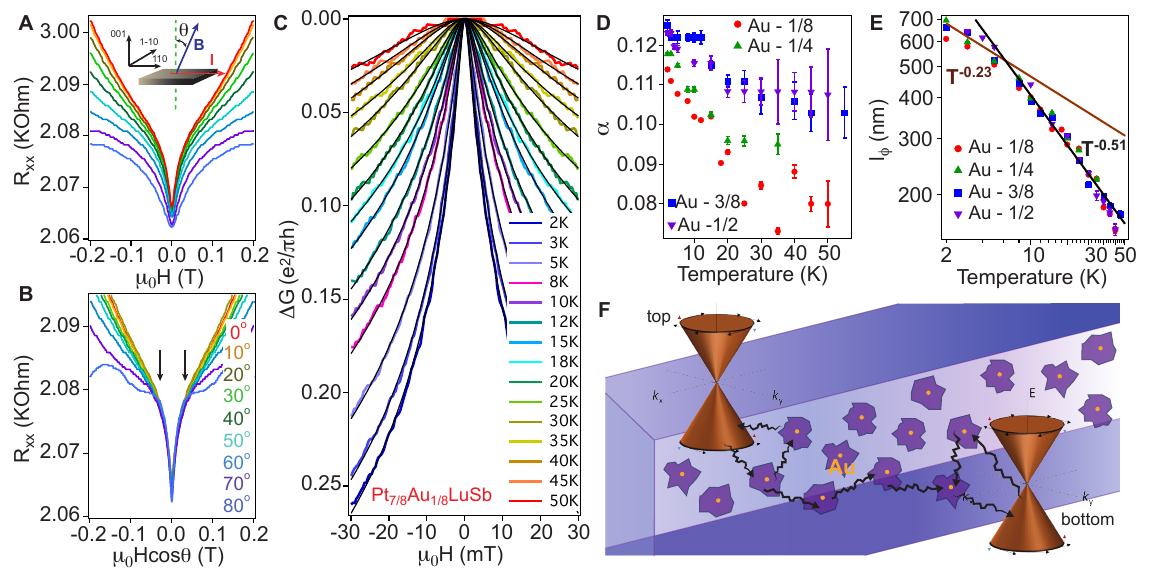}
\caption{Weak anti-localization in Pt$_{1-x}$Au$_{x}$LuSb thin films. (a) Weak-antilocalization (WAL) in Pt$_{7/8}$Au$_{1/8}$LuSb at different tilt angles. The tilt angle between the surface normal and the magnetic field vector is shown in the inset. (b) Same data in (a), but plotted as a function of the perpendicular component of the magnetic field vector. WAL behavior at low magnetic fields (the range shown by the black arrows) scales with the perpendicular component of the magnetic field vector indicating its two-dimensional character. Differential conductance $\Delta$G(B = $\mu_{0}$H) = G(B) - G(0) as a function of magnetic field and the corresponding HLN fits for (c) Pt$_{7/8}$Au$_{1/8}$LuSb. Temperature dependence of the extracted (d) pre-factor ($\alpha$) and (e) phase coherence length (l$_{\phi}$) as a function of temperature. (f) Cartoon showing interaction of the TSS with the quasi-localised bulk carriers. }
\label{fig:WAL}
\end{figure*}

\subsection{Weak anti-localization and evidence for surface-bulk coupling}
Weak anti-localization (WAL) observed at low magnetic fields in \pals\/ thin films, shown in Fig.~\ref{fig:WAL}, provides additional insights into their magnetoresistance behavior. Angle-dependent magnetotransport establishes two-dimensional nature of the observed WAL (Fig.~\ref{fig:WAL}(a)-(b)), ascribed to the presence of strongly spin-orbit coupled TSS as revealed by the ARPES measurements. WAL behavior in \pals\/ can be well described by the Hikami-Larkin-Nagaoka(HLN) theory \cite{hikami1980spin} given by
\begin{equation}
\Delta G = -\alpha\frac{e^{2}}{\pi h}[\Psi(\frac{1}{2} + \frac{B_{\phi}}{B_{\perp}}) - ln(\frac{B_{\phi}}{B_{\perp}})] ,
\end{equation}
where $\Psi$ is the digamma function, and $B_{\phi}  = \frac{\hbar}{4el_{\phi}^{2}}$ is the characteristic magnetic field corresponding to the phase coherence length $l_{\phi}$. The temperature dependence of the pre-factor $\alpha$ and the phase coherence length $l_{\phi}$ as a function of temperature, as extracted from the HLN fits, are shown in Fig.~\ref{fig:WAL}(d) and \ref{fig:WAL}(e), respectively. The magnitude of $\alpha$ is expected to be 0.5 for an independent strongly spin-orbit coupled two-dimensional channel \cite{hikami1980spin}. For typical topological insulators, $\alpha$ is expected to be 1 corresponding to two independent TSS at the opposite surfaces, which becomes equal to 0.5 when the two TSS at the opposite surfaces interact with each other leading to a single coherent channel \cite{ando2013topological}. $\alpha$ values of less than 0.5, as observed in the \pals\/ thin films, can arise due to the presence of strong disorder where higher order quantum corrections that goes beyond well-defined diffusive transport become important. In such cases, HLN analysis is still valid, but with a reduced pre-factor $\alpha'$ = (0.5 - 1/$\pi \gamma$), where $\gamma$ = $\frac{G}{e^{2}/h}$ \cite{liao2015observation, minkov2004magnetoresistance}. The obtained values of $\gamma$ in the range of 0.75-0.85 places \pals\/ thin films close to the weakly insulating regime (1$<$ $\gamma$ $<$ 3). It is observed that \pals\/ thin films with a lower Au concentration show enhanced tendency towards a stronger localization behavior ($\gamma$ $<<$ 1). The lower values of $\gamma$ corresponds to larger longitudinal resistance (R$_{xx}$, Fig.~\ref{fig:QH}) in \pals\/ thin films providing further evidence for a disorder driven reduction of the WAL effect. Reduced WAL effect ($\alpha$ $<$ 0.5) may also arise due to a reduction in the Berry phase $\phi = \pi(1- \Delta_{h}/2\mu)$ driven by a hybridization gap opening in the TSS ($\Delta_{h}$), when the Fermi level (E$_{F}$) is comparable to the hybridization gap \cite{kim2013coherent, lu2011competition}. Although, such an effect cannot be completely ruled out, we believe it plays a less significant role for the following reasons. First, the surface Fermi level changes by $\approx$ 150 meV (Fig.~\ref{fig:ARPES}(d)) between Pt$_{7/8}$Au$_{1/8}$LuSb and Pt$_{5/8}$Au$_{3/8}$LuSb, with it lying within the hybridization gap in Pt$_{5/8}$Au$_{3/8}$LuSb. In such a scenario, a much larger change in the $\alpha$ values would be expected between Pt$_{7/8}$Au$_{1/8}$LuSb and Pt$_{5/8}$Au$_{3/8}$LuSb, contrary to what has been experimentally observed. Second, a crossover from a WAL to a weak localization (WL) behavior is expected when the chemical potential lies within the hybridisation gap \cite{lu2011competition}, which is not observed in Pt$_{5/8}$Au$_{3/8}$LuSb. Furthermore, evidence for a strong disorder effect in \pals\/ thin films is also obtained by examining the temperature dependence of the phase coherence length ($\l_{\phi}$)  (Fig.~\ref{fig:WAL}(e)), which shows a power-law behavior $l_{\phi} \propto T^{-0.51}$. This is indicative of Nyquist dephasing due to electron-electron interaction effects in two-dimensional systems \cite{altshuler1982effects}. In addition, at low temperatures we find evidence for an enhanced dephasing rate, which is plausibly due to the coupling between the surface states and quasi-localised bulk carriers in the variable-range hopping (VRH) regime, as has been observed in other three-dimensional topological insulators \cite{liao2017enhanced}.

\section{Discussion}
\subsection{Two Dirac model}
\begin{figure}
\centering
\includegraphics[width=1\linewidth]{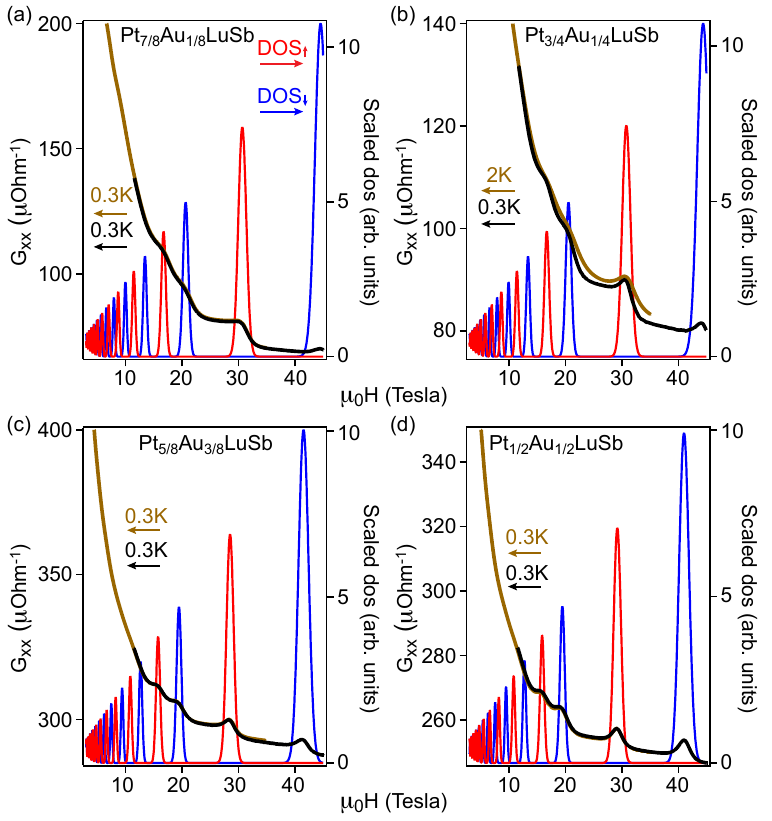}
\caption{Fits to the quantum Hall data in \pals\/. Conductance plots and corresponding fits to the \emph{two-Dirac model}, as described in the text, for (a) Pt$_{7/8}$Au$_{1/8}$LuSb, (b) Pt$_{3/4}$Au$_{1/4}$LuSb, (c) Pt$_{5/8}$Au$_{3/8}$LuSb, (d) Pt$_{1/2}$Au$_{1/2}$LuSb. DOS$_{\uparrow}$ and DOS$_{\downarrow}$ are the contributions from the Zeeman split Landau levels with spin directions along and opposite to the applied magnetic field, respectively.}
\label{fig:DiracModel}
\end{figure}

Having described the magnetotransport properties in our thin films we now construct a simple model to establish that the observed quantum Hall states originate from the topological surface states (TSS). We assume the presence of linearly dispersive TSS at both the surfaces of our thin films given by 
\begin{equation}
H_{TSS} = \hbar v_{f}(k_{x}\sigma_{y} - k_{y}\sigma_{x})\otimes \tau_{z} ,
\end{equation}
where the film surface vector is along the $z$ axis, $\sigma_i$ ($\tau_i$) represents the spin (surface) subspace\cite{brune2011quantum}. We further add a Zeeman term ($H_{Z}$), a hybridization term ($H_{hyb}$) that captures the interaction between the two TSS at opposite surfaces leading to the opening of a hybridization gap as observed in the ARPES measurements, and an inversion symmetry breaking term ($H_{invb}$) that lifts the degeneracy between the TSS at opposite surfaces, which could arise due to different screening effect and/or different surface potential at the two surfaces. The total Hamiltonian is

\begin{align}
\begin{split}\label{eq:1}
& H_{tot} = H_{TSS}+H_{Z}+H_{hyb}+H_{invb} ,\\
	& {\rm where}
\end{split}\\
& H_{Z} = g\mu_{B}B_{0}\sigma_{z}\otimes \mathds{1}\\
& H_{hyb} = \Delta_{h} \mathds{1} \otimes \tau_{x}\\
& H_{invb} = \Delta_{i} \mathds{1} \otimes \tau_{z}.
\end{align}

Note that the \textit{g}-factor is the effective \textit{g}-factor that incorporates orbital contributions of the bulk bands, thus capturing the modification of the surface state dispersion from the ideal Dirac-like behavior. Evidence for \textit{g}-factor anisotropy in the surface states is obtained from the angle-dependent magnetoresistance measurements (see Figs.~S8,S9 in \cite{suppl}) From the equations above we obtain the Landau energy levels given by
\begin{align}
\begin{split}\label{eq:2}
\mu = {}& \left[ 2v_{F}^{2}ne\hbar B_{0} + (g\mu_{B}B_{0})^{2} + \Delta_{h}^{2} + \Delta_{i}^{2} \right. \\
 & \left. \pm 2\sqrt{(g\mu_{B}B_{0}\Delta_{i})^{2}+(g\mu_{B}B_{0}\Delta_{h})^{2} + 2v_{F}^{2}ne\hbar B_{0}\Delta_{i}^{2}}\right]^{1/2} .
\end{split}
\end{align}
In Fig.~\ref{fig:DiracModel} we plot the calculated density of states from the above model, which describes the quantum Hall behavior in our thin films very well. We note that the Zeeman split Landau levels observed between 28 and 45 Tesla in \pals\/ thin films correspond to \emph{n} = 1 from where the carrier concentration of the TSS is obtained, which varies from 1.7$\times$10$^{12}$ cm$^{-2}$ in \pbls\/ to 1.8$\times$10$^{12}$ cm$^{-2}$ in \pdls\/. Inclusion of a finite g-factor and a hybridization gap of the TSS was found to be essential to model our data in agreement with the observation of a finite gap in TSS in the ARPES data. The extracted parameters from the model are summarised in Table~1. From magnetotransport, similar chemical potential values are obtained for two different Au concentrations viz. Pt$_{7/8}$Au$_{1/8}$LuSb and Pt$_{5/8}$Au$_{3/8}$LuSb at high magnetic fields (Table 1), which, however, differ significantly from the zero-field values obtained from the ARPES measurements (Fig.~\ref{fig:ARPES}, see $\Delta_{3}$ values in Table~S1 in \cite{suppl}). However, the chemical potential values in Pt$_{7/8}$Au$_{1/8}$LuSb, which has the lowest bulk carrier concentration, are found to be very similar at both zero and high magnetic fields. Such a behavior can be understood by noting that the quantum Hall states appear in \pals\/ thin films after they are in the bulk quantum limit \cite{suppl}, which results in a similar band-bending behavior at high magnetic fields for different gold concentrations, but can differ at low fields due to different bulk screening effect.  
\begin{table}
\centering
\caption{Model parameters}
\begin{ruledtabular}
\begin{tabular}{lrrrr}
Parameters & Au -1/8 & Au - 1/4 & Au - 3/8 & Au - 1/2\\
\hline
 1.  $\mu$ (meV) & 126 & 130 & 116 & 115  \\
 2. $g$ & 7.5 & 6.6 & 10.3 & 9.4\\
 3. $\Delta_{h}$ (meV) & 66 & 73 & 51 & 48\\
 4. $\Delta_{i}$ (meV) & 3 & 2 & 3 & 3 \\
 5. $k_{F}$ (\AA\/$^{-1}$) & 0.034 & 0.034 & 0.033 & 0.033\\

\end{tabular}
\end{ruledtabular}

\end{table}
\subsection{Ruling out alternate scenarios}
To rule out the possible bulk origin of the observed quantum Hall effect we explicitly calculated the Landau levels taking into account both the orbital and Zeeman effects along with the quantum confinement due to the finite thickness of our thin films. We utilized a \textit{k.p} fitting of the bulk-band structure along with the estimated Fermi levels from the low-field Hall coefficient (Fig.~\ref{fig:Transport}(d)) for the calculations, details of which can be found in the Supplementary Material \cite{suppl} (Section IV, Figs.~S10,S11). Even after the inclusion of a range of plausible g-factor values for the bulk bands, the experimental data cannot be reconciled with the estimated Landau levels ruling out the possibility of their bulk origin (see Fig.~S11 in \cite{suppl}). Another possible origin of the observed quantum Hall behavior is the presence of massive, doubly degenerate Volkov-Pankratov states at the \pals\/ / InSb hetero-interface \cite{volkov1985two, tchoumakov2017volkov, inhofer2017observation}. However, such states can only arise for a smooth hetero-interface where the gap evolves over a length \textit{l} greater than $\xi$, which is impossible in our case as the estimated value of $\xi$ in \pals\/ thin films exceeds the film thickness. Another tantalizing possibility for the origin of the observed quantum Hall effect could be the Fermi arc states \cite{armitage2018weyl} that originate from magnetic field induced Weyl points, as has been predicted in the half-Heusler compounds \cite{cano2017chiral}. However, symmetry constraints on the Fermi arc states explicitly rule out the possibility of lifting the degeneracy of the Fermi arc induced quantum Hall states under a magnetic field, in contrast to our experimental observation (see Figs.~S12,S13 in \cite{suppl}). 

We can also comprehensively rule out the possibility that the observed quantum Hall states might arise due to the surface accumulation layer in InSb, which is used as a buffer layer in \pals\/ thin films structures. First, as seen in Fig.~3, the quantum Hall states arise from p-type carriers as would be expected from TSS in \pals\/, whereas the carriers in InSb buffer layer are n-type and does not show any indication of a quantum Hall phase till 45 Tesla (See Fig.~S14 in \cite{suppl}). Second, the p-type carrier concentration estimated from the observed quantum Hall plateaus ranges from 1.7$\times$10$^{12}$ - 1.8$\times$10$^{12}$ cm$^{-2}$ in \pals\/ thin films. The only way this can arise from InSb would be if the surface Fermi level in InSb is pinned approximately 70 meV below the valence band maxima in InSb (see Section VI and Fig.~S16 in \cite{suppl}), which is extremely unlikely and to the best of our knowledge has never been reported in the literature. Indeed, direct measurement of the InSb (001) surface by scanning tunnelling spectroscopy (STS) reveals that the Fermi level is always pinned above the valence band maxima (see Section VI and Fig.~S15 in \cite{suppl}). Finally, another possibility could be bonding mismatch induced 2DEG at the \pals\//InSb heterointerface as has recently been observed for the case of LuSb (rock-salt)/GaSb (zinc-blende) structure\cite{chatterjee2021controlling}. However, in such cases the 2DEG is expected to have very low mobility and much higher carrier concentration than what is observed here. It is also noted that the half-Heusler crystal structure of \pals\/ contains a zinc-blende sub-lattice similar to InSb (zinc blende). Therefore, the bonding mismatch induced effects are expected to be much less pronounced at the \pals\//InSb heterointerface due to their higher degree of structural similarity, where a continuous Sb sub-lattice extends from InSb into \pals\/ layers (see Fig.~S1(d) in \cite{suppl}).

\subsection{Origin of quantum Hall effect and its evolution with Au concentration in \pals\/ thin films }
Our magnetotransport results, analysis of the WAL effect, and modelling of the quantum Hall data allows us to elucidate the origin of the linear magnetoresistance and its transmutation into a quantum Hall phase in \pals\/ thin films. At low Au concentration, TSS interact strongly with the semi-localized charge carriers in the bulk \cite{skinner2012bulk}, most likely formed around inhomogeneously distributed Au dopant atoms. This leads to a broadening of the Landau levels of the TSS, evident from the magnetoconductance data in Fig.~\ref{fig:QH}(e),(i), and manifests in a magnetoresistance behavior that is linear in magnetic field. Our data points to a mechanism similar to the one proposed by Wang and Lei \cite{wang2012linear}, where Zeeman splitting of the TSS plays an important role, evidence for which is provided in our analysis of the quantum Hall effect, shown in Fig.~\ref{fig:DiracModel}.  However, the predicted magnitude of the linear magnetoresistance within such a mechanism is much smaller than what we observe in \pals\/ thin films. It is possible that the experimentally observed strong coupling of the surface states with the quasi-localised bulk carriers enhances the LMR beyond what is predicted within the proposed mechanism \cite{nandi2018signatures}. In \pals\/ samples with higher Au concentration, bulk carriers exhibit stronger tendency to localize due to an increase in the impurity concentration, evident from a reduction in the bulk carrier mobility (Fig.~\ref{fig:Transport}(c)). This results in a weaker surface-bulk coupling resulting in the emergence of well-separated Landau levels of the TSS, which manifests as well defined quantum Hall plateaus. At sufficiently high fields, localization of the bulk carriers is enhanced and surface-bulk coupling is reduced even in the \pals\/ samples with low Au concentration, where a quantum Hall phase can be seen to emerge from a linear magnetoresistance behavior (Fig.~\ref{fig:QH}(a),(b)) at high magnetic field. Enhanced coupling between the surface and the bulk carriers in films with low Au concentration brings TSS closer to a strongly interacting regime, evident from a reduced WAL effect (Fig.~\ref{fig:WAL}(d)), as discussed earlier.

\section{Conclusion}

In summary, our work has revealed the macroscopic physical observables of the TSS in a semi-metallic half-Heusler compound \pals\/ in bulk measurements, thus establishing its topological nature and the potential of harnessing its topological properties in functional devices. We observe distinct plateaus in magnetoresistance in \pals\/ thin films, which we show arise from a TSS quantum Hall phase. This establishes compensation alloying in epitaxial thin films as an effective strategy to both reveal and control exotic properties in topological systems with a semi-metallic bulk band structure. Although, in the past, several different mechanisms have been suggested for the origin of linear magnetoresistance behavior in semimetals, our work experimentally establishes the role of TSS and its interaction with the quasi-localised bulk carriers in the observed linear magnetoresistance behavior by establishing it as a precursor phase to the quantum Hall phase emerging out of the TSS. This has been made possible by our ability to continuously tune the surface-bulk coupling within the same material system, which is shown as an important parameter controlling magnetotransport properties in compounds with both surface and bulk carriers.

\section*{Methods}

\subsection*{Thin Film Growth}

Epitaxial thin films of 15 nm thick \pals\/ were synthesized on a 270 nm thick InSb buffer layer grown on GaAs substrates by molecular-beam epitaxy (MBE). InSb atomic layers were nucleated at 360$^\circ$C and grown at 410$^\circ$C following thermal desorption of the native oxide on GaAs substrates. During the growth process, InSb (6.479 \AA) atomic layers were almost immediately relaxed due to a large lattice mismatch with GaAs (5.653 \AA). InSb buffer layers, which are grown in a modified VG-80 MBE system were taken out of the chamber post-growth and loaded into a modified Veeco Gen-II MBE system, where the surface oxide was removed by atomic hydrogen before the growth of \pals\/ atomic layers. \pals\/, which has a lattice constant similar to that of InSb, grows as a pseudomorphic epitaxial layer with respect to the underlying InSb buffer layer. \pals\/ thin films of different gold concentrations were all grown on the same InSb buffer layer, which was cleaved outside the vacuum chamber into smaller pieces for individual \pals\/ growths with different Au concentrations. PtLuSb thin films were grown at substrate temperatures between 370$^\circ$C and 400$^\circ$C. Substrate temperatures were progressively lowered for thin-film growths with higher gold concentrations with the Pt$_{1/2}$Au$_{1/2}$LuSb thin films grown at the substrate temperatures between 310$^\circ$C and 340$^\circ$C. Following a procedure similar to \cite{patel2014surface}, for \pals\/ growths Lu, Au, and Sb were evaporated from effusion cells, while Pt was evaporated using an e-beam evaporator. Atomic fluxes of Lu, Au, Sb, and Pt were calibrated by Rutherford back-scattering spectrometry (RBS) measurements of the elemental areal atomic density of calibration samples grown on Si substrates. These measurements were then used to calibrate \emph{in-situ} beam flux measurements using an ion gauge for Lu, Au, Sb, and a quartz crystal microbalance (QCM) for Pt.

\subsection*{ARPES}
ARPES measurements were performed at the Advanced Light Source at the end station 10.0.1.2. Samples were transferred from the growth chamber at UCSB to an analysis chamber equipped with a Scienta R4000 spectrometer at ALS in a custom-built UHV suitcase allowing us to maintain the pristine sample surface throughout the entire sample transfer process. Samples were cooled to a temperature of 70 K with liquid nitrogen. Tunable synchrotron light with photon energies in the range of 20 - 80 eV was used for the measurements.

\subsection*{Transport Measurements}
Transport measurements were performed on fabricated Hall bar devices using a low frequency ac lock-in technique. Measurements were performed in-house using a Quantum Design PPMS equipped with a 14 Tesla magnet and at the National High Magnetic Field Lab at Tallahassee, Florida using a 35 Tesla and a hybrid 45 Tesla magnets. Hall bars of dimension 400 $\mu$m $\times$ 100 $\mu$m were fabricated using optical lithography followed by an ion milling procedure using argon ions. Contacts were made using 50 $\mu$m gold wires bonded onto lithographically patterned gold contacts. A low current of 1 $\mu$A was sourced for all measurements to avoid Joule heating  with the current flowing along the [110] crystallographic axis, unless mentioned otherwise.

\subsection*{DFT Calculations}
The calculations were performed based on the DFT approach, as implemented in the VASP code \cite{VASP}. The exchange-correlation term was described using the generalized gradient approximation (GGA) functional proposed by Perdew, Burke, and Ernzerhof \cite{PBE}. The Kohn-Sham orbitals are expanded in a plane wave basis set with an energy cutoff of 400 eV. The Brillouin zone is sampled according to the Monkhorst-Pack method \cite{PhysRevB.13.5188}, using a gamma-centered $7\times7\times5$ ($7\times7\times1$) mesh for the tetragonal bulk (slab) calculations. The electron-ion interactions are taken into account using the projector augmented wave method \cite{PAW}. All geometries have been relaxed until atomic forces were lower than 0.01 eV/{\AA}. For the slab calculations, to avoid significant interaction between periodic images, a vacuum region of 15\,{\AA} was used.
\nocite{suppl,seibel2015gold,strohbeen2019electronically,dresselhaus1955spin,brydon2016pairing,potter2014quantum,davis1999evidence,liu1994surface,nextnano}

\section*{Author Contribution}S.C., J.A.L., and C.J.P. conceived the study. Thin film growth was performed by S.C. and J.A.L. Device fabrication and transport measurements were performed by S.C. with the assistance from D.R., A.G., C.D., H.I., Sha.C., and A.S. ARPES measurements were performed and analyzed by S.C. with the assistance from H.I., T.B., Y.C., D.R., and A.F. STM measurements were performed by N.W. and J.D. TEM measurements were performed by D.P. and A.G. DFT calculations were performed by F.C.L. and S.K. under the supervision of A.J. Theoretical analyses were performed by Y.F., S.C., and J.C. The manuscript was prepared by S.C. and C.J.P. All authors discussed results and commented on the manuscript.\\

\section*{Acknowledgements} 
We thank Mihir Pendharkar for helpful discussions. The synthesis of the thin films, the development of the UHV suitcase, the ARPES experiments, and the theoretical work were supported by the US Department of Energy (Contract no.~DE-SC0014388). Development of the growth facilities and low temperature magnetotransport measurements were supported by the Office of Naval Research through the Vannevar Bush Faculty Fellowship under the Award No. N00014-15-1-2845. Scanning probe studies were supported by the National Science Foundation (Award number DMR-1507875). This research used resources of the Advanced Light Source, which is a DOE Office of Science User Facility under contract no.~DE-AC02-05CH11231. A portion of this work was performed at the National High Magnetic Field Laboratory, which is supported by the National Science Foundation Cooperative Agreement DMR-1644779 and the state of Florida. We acknowledge the use of shared facilities of the National Science Foundation (NSF) Materials Research Science and Engineering Center (MRSEC) (DMR 1720256) at the University of California Santa Barbara, the UCSB Nanofabrication Facility, an open access laboratory, and the LeRoy Eyring Center for Solid State Science at Arizona State University. Density functional theory calculations made use of the National Energy Research Scientific Computing Center (NERSC), a U.S. Department of Energy Office of Science User Facility operated under Contract no.~DE-AC02-05CH11231. Theoretical analysis (J.C. and Y.F) was supported by the National Science Foundation under Grant No. DMR-1942447. J.C. acknowledges the support of the Flatiron Institute, a division of the Simons Foundation. D.R. gratefully acknowledges support from the Leverhulme Trust via an International Academic Fellowship (IAF-2018-039).


\begin{thebibliography}{66}%
\makeatletter
\providecommand \@ifxundefined [1]{%
 \@ifx{#1\undefined}
}%
\providecommand \@ifnum [1]{%
 \ifnum #1\expandafter \@firstoftwo
 \else \expandafter \@secondoftwo
 \fi
}%
\providecommand \@ifx [1]{%
 \ifx #1\expandafter \@firstoftwo
 \else \expandafter \@secondoftwo
 \fi
}%
\providecommand \natexlab [1]{#1}%
\providecommand \enquote  [1]{``#1''}%
\providecommand \bibnamefont  [1]{#1}%
\providecommand \bibfnamefont [1]{#1}%
\providecommand \citenamefont [1]{#1}%
\providecommand \href@noop [0]{\@secondoftwo}%
\providecommand \href [0]{\begingroup \@sanitize@url \@href}%
\providecommand \@href[1]{\@@startlink{#1}\@@href}%
\providecommand \@@href[1]{\endgroup#1\@@endlink}%
\providecommand \@sanitize@url [0]{\catcode `\\12\catcode `\$12\catcode
  `\&12\catcode `\#12\catcode `\^12\catcode `\_12\catcode `\%12\relax}%
\providecommand \@@startlink[1]{}%
\providecommand \@@endlink[0]{}%
\providecommand \url  [0]{\begingroup\@sanitize@url \@url }%
\providecommand \@url [1]{\endgroup\@href {#1}{\urlprefix }}%
\providecommand \urlprefix  [0]{URL }%
\providecommand \Eprint [0]{\href }%
\providecommand \doibase [0]{http://dx.doi.org/}%
\providecommand \selectlanguage [0]{\@gobble}%
\providecommand \bibinfo  [0]{\@secondoftwo}%
\providecommand \bibfield  [0]{\@secondoftwo}%
\providecommand \translation [1]{[#1]}%
\providecommand \BibitemOpen [0]{}%
\providecommand \bibitemStop [0]{}%
\providecommand \bibitemNoStop [0]{.\EOS\space}%
\providecommand \EOS [0]{\spacefactor3000\relax}%
\providecommand \BibitemShut  [1]{\csname bibitem#1\endcsname}%
\let\auto@bib@innerbib\@empty
\bibitem [{\citenamefont {Pippard}(1989)}]{pippard1989magnetoresistance}%
  \BibitemOpen
  \bibfield  {author} {\bibinfo {author} {\bibfnamefont {A.~B.}\ \bibnamefont
  {Pippard}},\ }\href@noop {} {\emph {\bibinfo {title} {Magnetoresistance in
  metals}}},\ Vol.~\bibinfo {volume} {2}\ (\bibinfo  {publisher} {Cambridge
  university press},\ \bibinfo {year} {1989})\BibitemShut {NoStop}%
\bibitem [{\citenamefont {Leahy}\ \emph {et~al.}(2018)\citenamefont {Leahy},
  \citenamefont {Lin}, \citenamefont {Siegfried}, \citenamefont {Treglia},
  \citenamefont {Song}, \citenamefont {Nandkishore},\ and\ \citenamefont
  {Lee}}]{leahy2018nonsaturating}%
  \BibitemOpen
  \bibfield  {author} {\bibinfo {author} {\bibfnamefont {I.~A.}\ \bibnamefont
  {Leahy}}, \bibinfo {author} {\bibfnamefont {Y.-P.}\ \bibnamefont {Lin}},
  \bibinfo {author} {\bibfnamefont {P.~E.}\ \bibnamefont {Siegfried}}, \bibinfo
  {author} {\bibfnamefont {A.~C.}\ \bibnamefont {Treglia}}, \bibinfo {author}
  {\bibfnamefont {J.~C.}\ \bibnamefont {Song}}, \bibinfo {author}
  {\bibfnamefont {R.~M.}\ \bibnamefont {Nandkishore}}, \ and\ \bibinfo {author}
  {\bibfnamefont {M.}~\bibnamefont {Lee}},\ }\href@noop {} {\bibfield
  {journal} {\bibinfo  {journal} {Proceedings of the National Academy of
  Sciences}\ }\textbf {\bibinfo {volume} {115}},\ \bibinfo {pages} {10570}
  (\bibinfo {year} {2018})}\BibitemShut {NoStop}%
\bibitem [{\citenamefont {Wang}\ \emph {et~al.}(2012)\citenamefont {Wang},
  \citenamefont {Du}, \citenamefont {Dou},\ and\ \citenamefont
  {Zhang}}]{wang2012room}%
  \BibitemOpen
  \bibfield  {author} {\bibinfo {author} {\bibfnamefont {X.}~\bibnamefont
  {Wang}}, \bibinfo {author} {\bibfnamefont {Y.}~\bibnamefont {Du}}, \bibinfo
  {author} {\bibfnamefont {S.}~\bibnamefont {Dou}}, \ and\ \bibinfo {author}
  {\bibfnamefont {C.}~\bibnamefont {Zhang}},\ }\href@noop {} {\bibfield
  {journal} {\bibinfo  {journal} {Physical Review Letters}\ }\textbf {\bibinfo
  {volume} {108}},\ \bibinfo {pages} {266806} (\bibinfo {year}
  {2012})}\BibitemShut {NoStop}%
\bibitem [{\citenamefont {Yan}\ \emph {et~al.}(2013)\citenamefont {Yan},
  \citenamefont {Wang}, \citenamefont {Yu},\ and\ \citenamefont
  {Liao}}]{yan2013large}%
  \BibitemOpen
  \bibfield  {author} {\bibinfo {author} {\bibfnamefont {Y.}~\bibnamefont
  {Yan}}, \bibinfo {author} {\bibfnamefont {L.-X.}\ \bibnamefont {Wang}},
  \bibinfo {author} {\bibfnamefont {D.-P.}\ \bibnamefont {Yu}}, \ and\ \bibinfo
  {author} {\bibfnamefont {Z.-M.}\ \bibnamefont {Liao}},\ }\href@noop {}
  {\bibfield  {journal} {\bibinfo  {journal} {Applied Physics Letters}\
  }\textbf {\bibinfo {volume} {103}},\ \bibinfo {pages} {033106} (\bibinfo
  {year} {2013})}\BibitemShut {NoStop}%
\bibitem [{\citenamefont {Zhang}\ \emph {et~al.}(2012)\citenamefont {Zhang},
  \citenamefont {McDonald}, \citenamefont {Shekhter}, \citenamefont {Bi},
  \citenamefont {Li}, \citenamefont {Jia},\ and\ \citenamefont
  {Picraux}}]{zhang2012magneto}%
  \BibitemOpen
  \bibfield  {author} {\bibinfo {author} {\bibfnamefont {S.}~\bibnamefont
  {Zhang}}, \bibinfo {author} {\bibfnamefont {R.}~\bibnamefont {McDonald}},
  \bibinfo {author} {\bibfnamefont {A.}~\bibnamefont {Shekhter}}, \bibinfo
  {author} {\bibfnamefont {Z.}~\bibnamefont {Bi}}, \bibinfo {author}
  {\bibfnamefont {Y.}~\bibnamefont {Li}}, \bibinfo {author} {\bibfnamefont
  {Q.}~\bibnamefont {Jia}}, \ and\ \bibinfo {author} {\bibfnamefont {S.~T.}\
  \bibnamefont {Picraux}},\ }\href@noop {} {\bibfield  {journal} {\bibinfo
  {journal} {Applied Physics Letters}\ }\textbf {\bibinfo {volume} {101}},\
  \bibinfo {pages} {202403} (\bibinfo {year} {2012})}\BibitemShut {NoStop}%
\bibitem [{\citenamefont {Narayanan}\ \emph {et~al.}(2015)\citenamefont
  {Narayanan}, \citenamefont {Watson}, \citenamefont {Blake}, \citenamefont
  {Bruyant}, \citenamefont {Drigo}, \citenamefont {Chen}, \citenamefont
  {Prabhakaran}, \citenamefont {Yan}, \citenamefont {Felser}, \citenamefont
  {Kong} \emph {et~al.}}]{narayanan2015linear}%
  \BibitemOpen
  \bibfield  {author} {\bibinfo {author} {\bibfnamefont {A.}~\bibnamefont
  {Narayanan}}, \bibinfo {author} {\bibfnamefont {M.}~\bibnamefont {Watson}},
  \bibinfo {author} {\bibfnamefont {S.}~\bibnamefont {Blake}}, \bibinfo
  {author} {\bibfnamefont {N.}~\bibnamefont {Bruyant}}, \bibinfo {author}
  {\bibfnamefont {L.}~\bibnamefont {Drigo}}, \bibinfo {author} {\bibfnamefont
  {Y.}~\bibnamefont {Chen}}, \bibinfo {author} {\bibfnamefont {D.}~\bibnamefont
  {Prabhakaran}}, \bibinfo {author} {\bibfnamefont {B.}~\bibnamefont {Yan}},
  \bibinfo {author} {\bibfnamefont {C.}~\bibnamefont {Felser}}, \bibinfo
  {author} {\bibfnamefont {T.}~\bibnamefont {Kong}},  \emph {et~al.},\
  }\href@noop {} {\bibfield  {journal} {\bibinfo  {journal} {Physical Review
  Letters}\ }\textbf {\bibinfo {volume} {114}},\ \bibinfo {pages} {117201}
  (\bibinfo {year} {2015})}\BibitemShut {NoStop}%
\bibitem [{\citenamefont {Xu}\ \emph {et~al.}(1997)\citenamefont {Xu},
  \citenamefont {Husmann}, \citenamefont {Rosenbaum}, \citenamefont {Saboungi},
  \citenamefont {Enderby},\ and\ \citenamefont {Littlewood}}]{xu1997large}%
  \BibitemOpen
  \bibfield  {author} {\bibinfo {author} {\bibfnamefont {R.}~\bibnamefont
  {Xu}}, \bibinfo {author} {\bibfnamefont {A.}~\bibnamefont {Husmann}},
  \bibinfo {author} {\bibfnamefont {T.}~\bibnamefont {Rosenbaum}}, \bibinfo
  {author} {\bibfnamefont {M.-L.}\ \bibnamefont {Saboungi}}, \bibinfo {author}
  {\bibfnamefont {J.}~\bibnamefont {Enderby}}, \ and\ \bibinfo {author}
  {\bibfnamefont {P.}~\bibnamefont {Littlewood}},\ }\href@noop {} {\bibfield
  {journal} {\bibinfo  {journal} {Nature}\ }\textbf {\bibinfo {volume} {390}},\
  \bibinfo {pages} {57} (\bibinfo {year} {1997})}\BibitemShut {NoStop}%
\bibitem [{\citenamefont {Parish}\ and\ \citenamefont
  {Littlewood}(2003)}]{parish2003non}%
  \BibitemOpen
  \bibfield  {author} {\bibinfo {author} {\bibfnamefont {M.}~\bibnamefont
  {Parish}}\ and\ \bibinfo {author} {\bibfnamefont {P.}~\bibnamefont
  {Littlewood}},\ }\href@noop {} {\bibfield  {journal} {\bibinfo  {journal}
  {Nature}\ }\textbf {\bibinfo {volume} {426}},\ \bibinfo {pages} {162}
  (\bibinfo {year} {2003})}\BibitemShut {NoStop}%
\bibitem [{\citenamefont {Song}\ \emph {et~al.}(2015)\citenamefont {Song},
  \citenamefont {Refael},\ and\ \citenamefont {Lee}}]{song2015linear}%
  \BibitemOpen
  \bibfield  {author} {\bibinfo {author} {\bibfnamefont {J.~C.}\ \bibnamefont
  {Song}}, \bibinfo {author} {\bibfnamefont {G.}~\bibnamefont {Refael}}, \ and\
  \bibinfo {author} {\bibfnamefont {P.~A.}\ \bibnamefont {Lee}},\ }\href@noop
  {} {\bibfield  {journal} {\bibinfo  {journal} {Physical Review B}\ }\textbf
  {\bibinfo {volume} {92}},\ \bibinfo {pages} {180204} (\bibinfo {year}
  {2015})}\BibitemShut {NoStop}%
\bibitem [{\citenamefont {Nandi}\ \emph {et~al.}(2018)\citenamefont {Nandi},
  \citenamefont {Skinner}, \citenamefont {Lee}, \citenamefont {Huang},
  \citenamefont {Shain}, \citenamefont {Chang}, \citenamefont {Ou},
  \citenamefont {Lee}, \citenamefont {Ward}, \citenamefont {Moodera} \emph
  {et~al.}}]{nandi2018signatures}%
  \BibitemOpen
  \bibfield  {author} {\bibinfo {author} {\bibfnamefont {D.}~\bibnamefont
  {Nandi}}, \bibinfo {author} {\bibfnamefont {B.}~\bibnamefont {Skinner}},
  \bibinfo {author} {\bibfnamefont {G.}~\bibnamefont {Lee}}, \bibinfo {author}
  {\bibfnamefont {K.-F.}\ \bibnamefont {Huang}}, \bibinfo {author}
  {\bibfnamefont {K.}~\bibnamefont {Shain}}, \bibinfo {author} {\bibfnamefont
  {C.-Z.}\ \bibnamefont {Chang}}, \bibinfo {author} {\bibfnamefont
  {Y.}~\bibnamefont {Ou}}, \bibinfo {author} {\bibfnamefont {S.-P.}\
  \bibnamefont {Lee}}, \bibinfo {author} {\bibfnamefont {J.}~\bibnamefont
  {Ward}}, \bibinfo {author} {\bibfnamefont {J.}~\bibnamefont {Moodera}},
  \emph {et~al.},\ }\href@noop {} {\bibfield  {journal} {\bibinfo  {journal}
  {Physical Review B}\ }\textbf {\bibinfo {volume} {98}},\ \bibinfo {pages}
  {214203} (\bibinfo {year} {2018})}\BibitemShut {NoStop}%
\bibitem [{\citenamefont {Abrikosov}(1998)}]{abrikosov1998quantum}%
  \BibitemOpen
  \bibfield  {author} {\bibinfo {author} {\bibfnamefont {A.}~\bibnamefont
  {Abrikosov}},\ }\href@noop {} {\bibfield  {journal} {\bibinfo  {journal}
  {Physical Review B}\ }\textbf {\bibinfo {volume} {58}},\ \bibinfo {pages}
  {2788} (\bibinfo {year} {1998})}\BibitemShut {NoStop}%
\bibitem [{\citenamefont {Wang}\ and\ \citenamefont
  {Lei}(2012)}]{wang2012linear}%
  \BibitemOpen
  \bibfield  {author} {\bibinfo {author} {\bibfnamefont {C.}~\bibnamefont
  {Wang}}\ and\ \bibinfo {author} {\bibfnamefont {X.}~\bibnamefont {Lei}},\
  }\href@noop {} {\bibfield  {journal} {\bibinfo  {journal} {Physical Review
  B}\ }\textbf {\bibinfo {volume} {86}},\ \bibinfo {pages} {035442} (\bibinfo
  {year} {2012})}\BibitemShut {NoStop}%
\bibitem [{\citenamefont {Assaf}\ \emph {et~al.}(2013)\citenamefont {Assaf},
  \citenamefont {Cardinal}, \citenamefont {Wei}, \citenamefont {Katmis},
  \citenamefont {Moodera},\ and\ \citenamefont {Heiman}}]{assaf2013linear}%
  \BibitemOpen
  \bibfield  {author} {\bibinfo {author} {\bibfnamefont {B.~A.}\ \bibnamefont
  {Assaf}}, \bibinfo {author} {\bibfnamefont {T.}~\bibnamefont {Cardinal}},
  \bibinfo {author} {\bibfnamefont {P.}~\bibnamefont {Wei}}, \bibinfo {author}
  {\bibfnamefont {F.}~\bibnamefont {Katmis}}, \bibinfo {author} {\bibfnamefont
  {J.~S.}\ \bibnamefont {Moodera}}, \ and\ \bibinfo {author} {\bibfnamefont
  {D.}~\bibnamefont {Heiman}},\ }\href@noop {} {\bibfield  {journal} {\bibinfo
  {journal} {Applied Physics Letters}\ }\textbf {\bibinfo {volume} {102}},\
  \bibinfo {pages} {012102} (\bibinfo {year} {2013})}\BibitemShut {NoStop}%
\bibitem [{\citenamefont {Burkov}(2016)}]{burkov2016topological}%
  \BibitemOpen
  \bibfield  {author} {\bibinfo {author} {\bibfnamefont {A.}~\bibnamefont
  {Burkov}},\ }\href@noop {} {\bibfield  {journal} {\bibinfo  {journal} {Nature
  Materials}\ }\textbf {\bibinfo {volume} {15}},\ \bibinfo {pages} {1145}
  (\bibinfo {year} {2016})}\BibitemShut {NoStop}%
\bibitem [{\citenamefont {Armitage}\ \emph {et~al.}(2018)\citenamefont
  {Armitage}, \citenamefont {Mele},\ and\ \citenamefont
  {Vishwanath}}]{armitage2018weyl}%
  \BibitemOpen
  \bibfield  {author} {\bibinfo {author} {\bibfnamefont {N.}~\bibnamefont
  {Armitage}}, \bibinfo {author} {\bibfnamefont {E.}~\bibnamefont {Mele}}, \
  and\ \bibinfo {author} {\bibfnamefont {A.}~\bibnamefont {Vishwanath}},\
  }\href@noop {} {\bibfield  {journal} {\bibinfo  {journal} {Reviews of Modern
  Physics}\ }\textbf {\bibinfo {volume} {90}},\ \bibinfo {pages} {015001}
  (\bibinfo {year} {2018})}\BibitemShut {NoStop}%
\bibitem [{\citenamefont {Klemenz}\ \emph {et~al.}(2019)\citenamefont
  {Klemenz}, \citenamefont {Lei},\ and\ \citenamefont
  {Schoop}}]{klemenz2019topological}%
  \BibitemOpen
  \bibfield  {author} {\bibinfo {author} {\bibfnamefont {S.}~\bibnamefont
  {Klemenz}}, \bibinfo {author} {\bibfnamefont {S.}~\bibnamefont {Lei}}, \ and\
  \bibinfo {author} {\bibfnamefont {L.~M.}\ \bibnamefont {Schoop}},\
  }\href@noop {} {\bibfield  {journal} {\bibinfo  {journal} {Annual Review of
  Materials Research}\ }\textbf {\bibinfo {volume} {49}},\ \bibinfo {pages}
  {185} (\bibinfo {year} {2019})}\BibitemShut {NoStop}%
\bibitem [{\citenamefont {Chadov}\ \emph {et~al.}(2010)\citenamefont {Chadov},
  \citenamefont {Qi}, \citenamefont {K{\"u}bler}, \citenamefont {Fecher},
  \citenamefont {Felser},\ and\ \citenamefont {Zhang}}]{chadov2010tunable}%
  \BibitemOpen
  \bibfield  {author} {\bibinfo {author} {\bibfnamefont {S.}~\bibnamefont
  {Chadov}}, \bibinfo {author} {\bibfnamefont {X.}~\bibnamefont {Qi}}, \bibinfo
  {author} {\bibfnamefont {J.}~\bibnamefont {K{\"u}bler}}, \bibinfo {author}
  {\bibfnamefont {G.~H.}\ \bibnamefont {Fecher}}, \bibinfo {author}
  {\bibfnamefont {C.}~\bibnamefont {Felser}}, \ and\ \bibinfo {author}
  {\bibfnamefont {S.~C.}\ \bibnamefont {Zhang}},\ }\href@noop {} {\bibfield
  {journal} {\bibinfo  {journal} {Nature Materials}\ }\textbf {\bibinfo
  {volume} {9}},\ \bibinfo {pages} {541} (\bibinfo {year} {2010})}\BibitemShut
  {NoStop}%
\bibitem [{\citenamefont {Lin}\ \emph {et~al.}(2010)\citenamefont {Lin},
  \citenamefont {Wray}, \citenamefont {Xia}, \citenamefont {Xu}, \citenamefont
  {Jia}, \citenamefont {Cava}, \citenamefont {Bansil},\ and\ \citenamefont
  {Hasan}}]{lin2010half}%
  \BibitemOpen
  \bibfield  {author} {\bibinfo {author} {\bibfnamefont {H.}~\bibnamefont
  {Lin}}, \bibinfo {author} {\bibfnamefont {L.~A.}\ \bibnamefont {Wray}},
  \bibinfo {author} {\bibfnamefont {Y.}~\bibnamefont {Xia}}, \bibinfo {author}
  {\bibfnamefont {S.}~\bibnamefont {Xu}}, \bibinfo {author} {\bibfnamefont
  {S.}~\bibnamefont {Jia}}, \bibinfo {author} {\bibfnamefont {R.~J.}\
  \bibnamefont {Cava}}, \bibinfo {author} {\bibfnamefont {A.}~\bibnamefont
  {Bansil}}, \ and\ \bibinfo {author} {\bibfnamefont {M.~Z.}\ \bibnamefont
  {Hasan}},\ }\href@noop {} {\bibfield  {journal} {\bibinfo  {journal} {Nature
  Materials}\ }\textbf {\bibinfo {volume} {9}},\ \bibinfo {pages} {546}
  (\bibinfo {year} {2010})}\BibitemShut {NoStop}%
\bibitem [{\citenamefont {Yang}\ \emph {et~al.}(2017)\citenamefont {Yang},
  \citenamefont {Yu}, \citenamefont {Parkin}, \citenamefont {Felser},
  \citenamefont {Liu},\ and\ \citenamefont {Yan}}]{yang2017prediction}%
  \BibitemOpen
  \bibfield  {author} {\bibinfo {author} {\bibfnamefont {H.}~\bibnamefont
  {Yang}}, \bibinfo {author} {\bibfnamefont {J.}~\bibnamefont {Yu}}, \bibinfo
  {author} {\bibfnamefont {S.~S.}\ \bibnamefont {Parkin}}, \bibinfo {author}
  {\bibfnamefont {C.}~\bibnamefont {Felser}}, \bibinfo {author} {\bibfnamefont
  {C.-X.}\ \bibnamefont {Liu}}, \ and\ \bibinfo {author} {\bibfnamefont
  {B.}~\bibnamefont {Yan}},\ }\href@noop {} {\bibfield  {journal} {\bibinfo
  {journal} {Physical Review Letters}\ }\textbf {\bibinfo {volume} {119}},\
  \bibinfo {pages} {136401} (\bibinfo {year} {2017})}\BibitemShut {NoStop}%
\bibitem [{\citenamefont {Wang}\ \emph {et~al.}(2016)\citenamefont {Wang},
  \citenamefont {Vergniory}, \citenamefont {Kushwaha}, \citenamefont
  {Hirschberger}, \citenamefont {Chulkov}, \citenamefont {Ernst}, \citenamefont
  {Ong}, \citenamefont {Cava},\ and\ \citenamefont {Bernevig}}]{wang2016time}%
  \BibitemOpen
  \bibfield  {author} {\bibinfo {author} {\bibfnamefont {Z.}~\bibnamefont
  {Wang}}, \bibinfo {author} {\bibfnamefont {M.}~\bibnamefont {Vergniory}},
  \bibinfo {author} {\bibfnamefont {S.}~\bibnamefont {Kushwaha}}, \bibinfo
  {author} {\bibfnamefont {M.}~\bibnamefont {Hirschberger}}, \bibinfo {author}
  {\bibfnamefont {E.}~\bibnamefont {Chulkov}}, \bibinfo {author} {\bibfnamefont
  {A.}~\bibnamefont {Ernst}}, \bibinfo {author} {\bibfnamefont {N.~P.}\
  \bibnamefont {Ong}}, \bibinfo {author} {\bibfnamefont {R.~J.}\ \bibnamefont
  {Cava}}, \ and\ \bibinfo {author} {\bibfnamefont {B.~A.}\ \bibnamefont
  {Bernevig}},\ }\href@noop {} {\bibfield  {journal} {\bibinfo  {journal}
  {Physical Review Letters}\ }\textbf {\bibinfo {volume} {117}},\ \bibinfo
  {pages} {236401} (\bibinfo {year} {2016})}\BibitemShut {NoStop}%
\bibitem [{\citenamefont {Chang}\ \emph {et~al.}(2017)\citenamefont {Chang},
  \citenamefont {Xu}, \citenamefont {Zhou}, \citenamefont {Huang},
  \citenamefont {Singh}, \citenamefont {Wang}, \citenamefont {Belopolski},
  \citenamefont {Yin}, \citenamefont {Zhang}, \citenamefont {Bansil} \emph
  {et~al.}}]{chang2017topological}%
  \BibitemOpen
  \bibfield  {author} {\bibinfo {author} {\bibfnamefont {G.}~\bibnamefont
  {Chang}}, \bibinfo {author} {\bibfnamefont {S.-Y.}\ \bibnamefont {Xu}},
  \bibinfo {author} {\bibfnamefont {X.}~\bibnamefont {Zhou}}, \bibinfo {author}
  {\bibfnamefont {S.-M.}\ \bibnamefont {Huang}}, \bibinfo {author}
  {\bibfnamefont {B.}~\bibnamefont {Singh}}, \bibinfo {author} {\bibfnamefont
  {B.}~\bibnamefont {Wang}}, \bibinfo {author} {\bibfnamefont {I.}~\bibnamefont
  {Belopolski}}, \bibinfo {author} {\bibfnamefont {J.}~\bibnamefont {Yin}},
  \bibinfo {author} {\bibfnamefont {S.}~\bibnamefont {Zhang}}, \bibinfo
  {author} {\bibfnamefont {A.}~\bibnamefont {Bansil}},  \emph {et~al.},\
  }\href@noop {} {\bibfield  {journal} {\bibinfo  {journal} {Physical Review
  Letters}\ }\textbf {\bibinfo {volume} {119}},\ \bibinfo {pages} {156401}
  (\bibinfo {year} {2017})}\BibitemShut {NoStop}%
\bibitem [{\citenamefont {Chang}\ \emph {et~al.}(2016)\citenamefont {Chang},
  \citenamefont {Xu}, \citenamefont {Zheng}, \citenamefont {Singh},
  \citenamefont {Hsu}, \citenamefont {Bian}, \citenamefont {Alidoust},
  \citenamefont {Belopolski}, \citenamefont {Sanchez}, \citenamefont {Zhang}
  \emph {et~al.}}]{chang2016room}%
  \BibitemOpen
  \bibfield  {author} {\bibinfo {author} {\bibfnamefont {G.}~\bibnamefont
  {Chang}}, \bibinfo {author} {\bibfnamefont {S.-Y.}\ \bibnamefont {Xu}},
  \bibinfo {author} {\bibfnamefont {H.}~\bibnamefont {Zheng}}, \bibinfo
  {author} {\bibfnamefont {B.}~\bibnamefont {Singh}}, \bibinfo {author}
  {\bibfnamefont {C.-H.}\ \bibnamefont {Hsu}}, \bibinfo {author} {\bibfnamefont
  {G.}~\bibnamefont {Bian}}, \bibinfo {author} {\bibfnamefont {N.}~\bibnamefont
  {Alidoust}}, \bibinfo {author} {\bibfnamefont {I.}~\bibnamefont
  {Belopolski}}, \bibinfo {author} {\bibfnamefont {D.~S.}\ \bibnamefont
  {Sanchez}}, \bibinfo {author} {\bibfnamefont {S.}~\bibnamefont {Zhang}},
  \emph {et~al.},\ }\href@noop {} {\bibfield  {journal} {\bibinfo  {journal}
  {Scientific Reports}\ }\textbf {\bibinfo {volume} {6}},\ \bibinfo {pages}
  {38839} (\bibinfo {year} {2016})}\BibitemShut {NoStop}%
\bibitem [{\citenamefont {Logan}\ \emph {et~al.}(2016)\citenamefont {Logan},
  \citenamefont {Patel}, \citenamefont {Harrington}, \citenamefont {Polley},
  \citenamefont {Schultz}, \citenamefont {Balasubramanian}, \citenamefont
  {Janotti}, \citenamefont {Mikkelsen},\ and\ \citenamefont
  {Palmstr{\o}m}}]{logan2016observation}%
  \BibitemOpen
  \bibfield  {author} {\bibinfo {author} {\bibfnamefont {J.~A.}\ \bibnamefont
  {Logan}}, \bibinfo {author} {\bibfnamefont {S.}~\bibnamefont {Patel}},
  \bibinfo {author} {\bibfnamefont {S.~D.}\ \bibnamefont {Harrington}},
  \bibinfo {author} {\bibfnamefont {C.}~\bibnamefont {Polley}}, \bibinfo
  {author} {\bibfnamefont {B.~D.}\ \bibnamefont {Schultz}}, \bibinfo {author}
  {\bibfnamefont {T.}~\bibnamefont {Balasubramanian}}, \bibinfo {author}
  {\bibfnamefont {A.}~\bibnamefont {Janotti}}, \bibinfo {author} {\bibfnamefont
  {A.}~\bibnamefont {Mikkelsen}}, \ and\ \bibinfo {author} {\bibfnamefont
  {C.~J.}\ \bibnamefont {Palmstr{\o}m}},\ }\href@noop {} {\bibfield  {journal}
  {\bibinfo  {journal} {Nature Communications}\ }\textbf {\bibinfo {volume}
  {7}},\ \bibinfo {pages} {11993} (\bibinfo {year} {2016})}\BibitemShut
  {NoStop}%
\bibitem [{\citenamefont {Yonggang}\ \emph {et~al.}(2017)\citenamefont
  {Yonggang}, \citenamefont {Zhang},\ and\ \citenamefont
  {Zunger}}]{yonggang2017natural}%
  \BibitemOpen
  \bibfield  {author} {\bibinfo {author} {\bibfnamefont {G.~Y.}\ \bibnamefont
  {Yonggang}}, \bibinfo {author} {\bibfnamefont {X.}~\bibnamefont {Zhang}}, \
  and\ \bibinfo {author} {\bibfnamefont {A.}~\bibnamefont {Zunger}},\
  }\href@noop {} {\bibfield  {journal} {\bibinfo  {journal} {Physical Review
  B}\ }\textbf {\bibinfo {volume} {95}},\ \bibinfo {pages} {085201} (\bibinfo
  {year} {2017})}\BibitemShut {NoStop}%
\bibitem [{\citenamefont {Patel}\ \emph {et~al.}(2014)\citenamefont {Patel},
  \citenamefont {Kawasaki}, \citenamefont {Logan}, \citenamefont {Schultz},
  \citenamefont {Adell}, \citenamefont {Thiagarajan}, \citenamefont
  {Mikkelsen},\ and\ \citenamefont {Palmstr{\o}m}}]{patel2014surface}%
  \BibitemOpen
  \bibfield  {author} {\bibinfo {author} {\bibfnamefont {S.~J.}\ \bibnamefont
  {Patel}}, \bibinfo {author} {\bibfnamefont {J.~K.}\ \bibnamefont {Kawasaki}},
  \bibinfo {author} {\bibfnamefont {J.}~\bibnamefont {Logan}}, \bibinfo
  {author} {\bibfnamefont {B.~D.}\ \bibnamefont {Schultz}}, \bibinfo {author}
  {\bibfnamefont {J.}~\bibnamefont {Adell}}, \bibinfo {author} {\bibfnamefont
  {B.}~\bibnamefont {Thiagarajan}}, \bibinfo {author} {\bibfnamefont
  {A.}~\bibnamefont {Mikkelsen}}, \ and\ \bibinfo {author} {\bibfnamefont
  {C.~J.}\ \bibnamefont {Palmstr{\o}m}},\ }\href@noop {} {\bibfield  {journal}
  {\bibinfo  {journal} {Applied Physics Letters}\ }\textbf {\bibinfo {volume}
  {104}},\ \bibinfo {pages} {201603} (\bibinfo {year} {2014})}\BibitemShut
  {NoStop}%
\bibitem [{\citenamefont {Mellnik}\ \emph {et~al.}(2014)\citenamefont
  {Mellnik}, \citenamefont {Lee}, \citenamefont {Richardella}, \citenamefont
  {Grab}, \citenamefont {Mintun}, \citenamefont {Fischer}, \citenamefont
  {Vaezi}, \citenamefont {Manchon}, \citenamefont {Kim}, \citenamefont
  {Samarth} \emph {et~al.}}]{mellnik2014spin}%
  \BibitemOpen
  \bibfield  {author} {\bibinfo {author} {\bibfnamefont {A.}~\bibnamefont
  {Mellnik}}, \bibinfo {author} {\bibfnamefont {J.}~\bibnamefont {Lee}},
  \bibinfo {author} {\bibfnamefont {A.}~\bibnamefont {Richardella}}, \bibinfo
  {author} {\bibfnamefont {J.}~\bibnamefont {Grab}}, \bibinfo {author}
  {\bibfnamefont {P.}~\bibnamefont {Mintun}}, \bibinfo {author} {\bibfnamefont
  {M.~H.}\ \bibnamefont {Fischer}}, \bibinfo {author} {\bibfnamefont
  {A.}~\bibnamefont {Vaezi}}, \bibinfo {author} {\bibfnamefont
  {A.}~\bibnamefont {Manchon}}, \bibinfo {author} {\bibfnamefont {E.-A.}\
  \bibnamefont {Kim}}, \bibinfo {author} {\bibfnamefont {N.}~\bibnamefont
  {Samarth}},  \emph {et~al.},\ }\href@noop {} {\bibfield  {journal} {\bibinfo
  {journal} {Nature}\ }\textbf {\bibinfo {volume} {511}},\ \bibinfo {pages}
  {449} (\bibinfo {year} {2014})}\BibitemShut {NoStop}%
\bibitem [{\citenamefont {Qi}\ and\ \citenamefont
  {Zhang}(2011)}]{qi2011topological}%
  \BibitemOpen
  \bibfield  {author} {\bibinfo {author} {\bibfnamefont {X.-L.}\ \bibnamefont
  {Qi}}\ and\ \bibinfo {author} {\bibfnamefont {S.-C.}\ \bibnamefont {Zhang}},\
  }\href@noop {} {\bibfield  {journal} {\bibinfo  {journal} {Reviews of Modern
  Physics}\ }\textbf {\bibinfo {volume} {83}},\ \bibinfo {pages} {1057}
  (\bibinfo {year} {2011})}\BibitemShut {NoStop}%
\bibitem [{\citenamefont {De~Groot}\ \emph {et~al.}(1983)\citenamefont
  {De~Groot}, \citenamefont {Mueller}, \citenamefont {Van~Engen},\ and\
  \citenamefont {Buschow}}]{de1983new}%
  \BibitemOpen
  \bibfield  {author} {\bibinfo {author} {\bibfnamefont {R.}~\bibnamefont
  {De~Groot}}, \bibinfo {author} {\bibfnamefont {F.}~\bibnamefont {Mueller}},
  \bibinfo {author} {\bibfnamefont {P.}~\bibnamefont {Van~Engen}}, \ and\
  \bibinfo {author} {\bibfnamefont {K.}~\bibnamefont {Buschow}},\ }\href@noop
  {} {\bibfield  {journal} {\bibinfo  {journal} {Physical Review Letters}\
  }\textbf {\bibinfo {volume} {50}},\ \bibinfo {pages} {2024} (\bibinfo {year}
  {1983})}\BibitemShut {NoStop}%
\bibitem [{\citenamefont {Klimczuk}\ \emph {et~al.}(2012)\citenamefont
  {Klimczuk}, \citenamefont {Wang}, \citenamefont {Gofryk}, \citenamefont
  {Ronning}, \citenamefont {Winterlik}, \citenamefont {Fecher}, \citenamefont
  {Griveau}, \citenamefont {Colineau}, \citenamefont {Felser}, \citenamefont
  {Thompson} \emph {et~al.}}]{klimczuk2012superconductivity}%
  \BibitemOpen
  \bibfield  {author} {\bibinfo {author} {\bibfnamefont {T.}~\bibnamefont
  {Klimczuk}}, \bibinfo {author} {\bibfnamefont {C.}~\bibnamefont {Wang}},
  \bibinfo {author} {\bibfnamefont {K.}~\bibnamefont {Gofryk}}, \bibinfo
  {author} {\bibfnamefont {F.}~\bibnamefont {Ronning}}, \bibinfo {author}
  {\bibfnamefont {J.}~\bibnamefont {Winterlik}}, \bibinfo {author}
  {\bibfnamefont {G.}~\bibnamefont {Fecher}}, \bibinfo {author} {\bibfnamefont
  {J.-C.}\ \bibnamefont {Griveau}}, \bibinfo {author} {\bibfnamefont
  {E.}~\bibnamefont {Colineau}}, \bibinfo {author} {\bibfnamefont
  {C.}~\bibnamefont {Felser}}, \bibinfo {author} {\bibfnamefont {J.~D.}\
  \bibnamefont {Thompson}},  \emph {et~al.},\ }\href@noop {} {\bibfield
  {journal} {\bibinfo  {journal} {Physical Review B}\ }\textbf {\bibinfo
  {volume} {85}},\ \bibinfo {pages} {174505} (\bibinfo {year}
  {2012})}\BibitemShut {NoStop}%
\bibitem [{\citenamefont {Nakajima}\ \emph {et~al.}(2015)\citenamefont
  {Nakajima}, \citenamefont {Hu}, \citenamefont {Kirshenbaum}, \citenamefont
  {Hughes}, \citenamefont {Syers}, \citenamefont {Wang}, \citenamefont {Wang},
  \citenamefont {Wang}, \citenamefont {Saha}, \citenamefont {Pratt} \emph
  {et~al.}}]{nakajima2015topological}%
  \BibitemOpen
  \bibfield  {author} {\bibinfo {author} {\bibfnamefont {Y.}~\bibnamefont
  {Nakajima}}, \bibinfo {author} {\bibfnamefont {R.}~\bibnamefont {Hu}},
  \bibinfo {author} {\bibfnamefont {K.}~\bibnamefont {Kirshenbaum}}, \bibinfo
  {author} {\bibfnamefont {A.}~\bibnamefont {Hughes}}, \bibinfo {author}
  {\bibfnamefont {P.}~\bibnamefont {Syers}}, \bibinfo {author} {\bibfnamefont
  {X.}~\bibnamefont {Wang}}, \bibinfo {author} {\bibfnamefont {K.}~\bibnamefont
  {Wang}}, \bibinfo {author} {\bibfnamefont {R.}~\bibnamefont {Wang}}, \bibinfo
  {author} {\bibfnamefont {S.~R.}\ \bibnamefont {Saha}}, \bibinfo {author}
  {\bibfnamefont {D.}~\bibnamefont {Pratt}},  \emph {et~al.},\ }\href@noop {}
  {\bibfield  {journal} {\bibinfo  {journal} {Science Advances}\ }\textbf
  {\bibinfo {volume} {1}},\ \bibinfo {pages} {e1500242} (\bibinfo {year}
  {2015})}\BibitemShut {NoStop}%
\bibitem [{\citenamefont {Hinterleitner}\ \emph {et~al.}(2019)\citenamefont
  {Hinterleitner}, \citenamefont {Knapp}, \citenamefont {Poneder},
  \citenamefont {Shi}, \citenamefont {M{\"u}ller}, \citenamefont {Eguchi},
  \citenamefont {Eisenmenger-Sittner}, \citenamefont {St{\"o}ger-Pollach},
  \citenamefont {Kakefuda}, \citenamefont {Kawamoto} \emph
  {et~al.}}]{hinterleitner2019thermoelectric}%
  \BibitemOpen
  \bibfield  {author} {\bibinfo {author} {\bibfnamefont {B.}~\bibnamefont
  {Hinterleitner}}, \bibinfo {author} {\bibfnamefont {I.}~\bibnamefont
  {Knapp}}, \bibinfo {author} {\bibfnamefont {M.}~\bibnamefont {Poneder}},
  \bibinfo {author} {\bibfnamefont {Y.}~\bibnamefont {Shi}}, \bibinfo {author}
  {\bibfnamefont {H.}~\bibnamefont {M{\"u}ller}}, \bibinfo {author}
  {\bibfnamefont {G.}~\bibnamefont {Eguchi}}, \bibinfo {author} {\bibfnamefont
  {C.}~\bibnamefont {Eisenmenger-Sittner}}, \bibinfo {author} {\bibfnamefont
  {M.}~\bibnamefont {St{\"o}ger-Pollach}}, \bibinfo {author} {\bibfnamefont
  {Y.}~\bibnamefont {Kakefuda}}, \bibinfo {author} {\bibfnamefont
  {N.}~\bibnamefont {Kawamoto}},  \emph {et~al.},\ }\href@noop {} {\bibfield
  {journal} {\bibinfo  {journal} {Nature}\ }\textbf {\bibinfo {volume} {576}},\
  \bibinfo {pages} {85} (\bibinfo {year} {2019})}\BibitemShut {NoStop}%
\bibitem [{\citenamefont {Wollmann}\ \emph {et~al.}(2017)\citenamefont
  {Wollmann}, \citenamefont {Nayak}, \citenamefont {Parkin},\ and\
  \citenamefont {Felser}}]{wollmann2017heusler}%
  \BibitemOpen
  \bibfield  {author} {\bibinfo {author} {\bibfnamefont {L.}~\bibnamefont
  {Wollmann}}, \bibinfo {author} {\bibfnamefont {A.~K.}\ \bibnamefont {Nayak}},
  \bibinfo {author} {\bibfnamefont {S.~S.}\ \bibnamefont {Parkin}}, \ and\
  \bibinfo {author} {\bibfnamefont {C.}~\bibnamefont {Felser}},\ }\href@noop {}
  {\bibfield  {journal} {\bibinfo  {journal} {Annual Review of Materials
  Research}\ }\textbf {\bibinfo {volume} {47}},\ \bibinfo {pages} {247}
  (\bibinfo {year} {2017})}\BibitemShut {NoStop}%
\bibitem [{\citenamefont {Graf}\ \emph {et~al.}(2011)\citenamefont {Graf},
  \citenamefont {Felser},\ and\ \citenamefont {Parkin}}]{graf2011simple}%
  \BibitemOpen
  \bibfield  {author} {\bibinfo {author} {\bibfnamefont {T.}~\bibnamefont
  {Graf}}, \bibinfo {author} {\bibfnamefont {C.}~\bibnamefont {Felser}}, \ and\
  \bibinfo {author} {\bibfnamefont {S.~S.}\ \bibnamefont {Parkin}},\
  }\href@noop {} {\bibfield  {journal} {\bibinfo  {journal} {Progress in Solid
  State Chemistry}\ }\textbf {\bibinfo {volume} {39}},\ \bibinfo {pages} {1}
  (\bibinfo {year} {2011})}\BibitemShut {NoStop}%
\bibitem [{\citenamefont {Palmstr{\o}m}(2016)}]{palmstrom2016heusler}%
  \BibitemOpen
  \bibfield  {author} {\bibinfo {author} {\bibfnamefont {C.~J.}\ \bibnamefont
  {Palmstr{\o}m}},\ }\href@noop {} {\bibfield  {journal} {\bibinfo  {journal}
  {Progress in Crystal Growth and Characterization of Materials}\ }\textbf
  {\bibinfo {volume} {62}},\ \bibinfo {pages} {371} (\bibinfo {year}
  {2016})}\BibitemShut {NoStop}%
\bibitem [{sup()}]{suppl}%
  \BibitemOpen
  \href@noop {} {}\bibinfo {note} {See Supplemental Material [url], which
  includes Refs. [59-66]}\BibitemShut {NoStop}%
\bibitem [{\citenamefont {Linder}\ \emph {et~al.}(2009)\citenamefont {Linder},
  \citenamefont {Yokoyama},\ and\ \citenamefont
  {Sudb{\o}}}]{linder2009anomalous}%
  \BibitemOpen
  \bibfield  {author} {\bibinfo {author} {\bibfnamefont {J.}~\bibnamefont
  {Linder}}, \bibinfo {author} {\bibfnamefont {T.}~\bibnamefont {Yokoyama}}, \
  and\ \bibinfo {author} {\bibfnamefont {A.}~\bibnamefont {Sudb{\o}}},\
  }\href@noop {} {\bibfield  {journal} {\bibinfo  {journal} {Physical Review
  B}\ }\textbf {\bibinfo {volume} {80}},\ \bibinfo {pages} {205401} (\bibinfo
  {year} {2009})}\BibitemShut {NoStop}%
\bibitem [{\citenamefont {Brillson}(2010)}]{brillson2010surfaces}%
  \BibitemOpen
  \bibfield  {author} {\bibinfo {author} {\bibfnamefont {L.~J.}\ \bibnamefont
  {Brillson}},\ }\href@noop {} {\emph {\bibinfo {title} {Surfaces and
  interfaces of electronic materials}}}\ (\bibinfo  {publisher} {John Wiley \&
  Sons},\ \bibinfo {year} {2010})\BibitemShut {NoStop}%
\bibitem [{\citenamefont {Singh}\ \emph {et~al.}(2017)\citenamefont {Singh},
  \citenamefont {Gopal}, \citenamefont {Sarkar}, \citenamefont {Pandey},
  \citenamefont {Patel},\ and\ \citenamefont {Mitra}}]{singh2017linear}%
  \BibitemOpen
  \bibfield  {author} {\bibinfo {author} {\bibfnamefont {S.}~\bibnamefont
  {Singh}}, \bibinfo {author} {\bibfnamefont {R.}~\bibnamefont {Gopal}},
  \bibinfo {author} {\bibfnamefont {J.}~\bibnamefont {Sarkar}}, \bibinfo
  {author} {\bibfnamefont {A.}~\bibnamefont {Pandey}}, \bibinfo {author}
  {\bibfnamefont {B.~G.}\ \bibnamefont {Patel}}, \ and\ \bibinfo {author}
  {\bibfnamefont {C.}~\bibnamefont {Mitra}},\ }\href@noop {} {\bibfield
  {journal} {\bibinfo  {journal} {Journal of Physics: Condensed Matter}\
  }\textbf {\bibinfo {volume} {29}},\ \bibinfo {pages} {505601} (\bibinfo
  {year} {2017})}\BibitemShut {NoStop}%
\bibitem [{\citenamefont {Ando}(2013)}]{ando2013topological}%
  \BibitemOpen
  \bibfield  {author} {\bibinfo {author} {\bibfnamefont {Y.}~\bibnamefont
  {Ando}},\ }\href@noop {} {\bibfield  {journal} {\bibinfo  {journal} {Journal
  of the Physical Society of Japan}\ }\textbf {\bibinfo {volume} {82}},\
  \bibinfo {pages} {102001} (\bibinfo {year} {2013})}\BibitemShut {NoStop}%
\bibitem [{\citenamefont {Lei}\ \emph {et~al.}(2019)\citenamefont {Lei},
  \citenamefont {Lehner}, \citenamefont {Cheah}, \citenamefont {Karalic},
  \citenamefont {Mittag}, \citenamefont {Alt}, \citenamefont {Scharnetzky},
  \citenamefont {Wegscheider}, \citenamefont {Ihn},\ and\ \citenamefont
  {Ensslin}}]{lei2019quantum}%
  \BibitemOpen
  \bibfield  {author} {\bibinfo {author} {\bibfnamefont {Z.}~\bibnamefont
  {Lei}}, \bibinfo {author} {\bibfnamefont {C.~A.}\ \bibnamefont {Lehner}},
  \bibinfo {author} {\bibfnamefont {E.}~\bibnamefont {Cheah}}, \bibinfo
  {author} {\bibfnamefont {M.}~\bibnamefont {Karalic}}, \bibinfo {author}
  {\bibfnamefont {C.}~\bibnamefont {Mittag}}, \bibinfo {author} {\bibfnamefont
  {L.}~\bibnamefont {Alt}}, \bibinfo {author} {\bibfnamefont {J.}~\bibnamefont
  {Scharnetzky}}, \bibinfo {author} {\bibfnamefont {W.}~\bibnamefont
  {Wegscheider}}, \bibinfo {author} {\bibfnamefont {T.}~\bibnamefont {Ihn}}, \
  and\ \bibinfo {author} {\bibfnamefont {K.}~\bibnamefont {Ensslin}},\
  }\href@noop {} {\bibfield  {journal} {\bibinfo  {journal} {Applied Physics
  Letters}\ }\textbf {\bibinfo {volume} {115}},\ \bibinfo {pages} {012101}
  (\bibinfo {year} {2019})}\BibitemShut {NoStop}%
\bibitem [{\citenamefont {Hikami}\ \emph {et~al.}(1980)\citenamefont {Hikami},
  \citenamefont {Larkin},\ and\ \citenamefont {Nagaoka}}]{hikami1980spin}%
  \BibitemOpen
  \bibfield  {author} {\bibinfo {author} {\bibfnamefont {S.}~\bibnamefont
  {Hikami}}, \bibinfo {author} {\bibfnamefont {A.~I.}\ \bibnamefont {Larkin}},
  \ and\ \bibinfo {author} {\bibfnamefont {Y.}~\bibnamefont {Nagaoka}},\
  }\href@noop {} {\bibfield  {journal} {\bibinfo  {journal} {Progress of
  Theoretical Physics}\ }\textbf {\bibinfo {volume} {63}},\ \bibinfo {pages}
  {707} (\bibinfo {year} {1980})}\BibitemShut {NoStop}%
\bibitem [{\citenamefont {Liao}\ \emph {et~al.}(2015)\citenamefont {Liao},
  \citenamefont {Ou}, \citenamefont {Feng}, \citenamefont {Yang}, \citenamefont
  {Lin}, \citenamefont {Yang}, \citenamefont {Wu}, \citenamefont {He},
  \citenamefont {Ma}, \citenamefont {Xue} \emph
  {et~al.}}]{liao2015observation}%
  \BibitemOpen
  \bibfield  {author} {\bibinfo {author} {\bibfnamefont {J.}~\bibnamefont
  {Liao}}, \bibinfo {author} {\bibfnamefont {Y.}~\bibnamefont {Ou}}, \bibinfo
  {author} {\bibfnamefont {X.}~\bibnamefont {Feng}}, \bibinfo {author}
  {\bibfnamefont {S.}~\bibnamefont {Yang}}, \bibinfo {author} {\bibfnamefont
  {C.}~\bibnamefont {Lin}}, \bibinfo {author} {\bibfnamefont {W.}~\bibnamefont
  {Yang}}, \bibinfo {author} {\bibfnamefont {K.}~\bibnamefont {Wu}}, \bibinfo
  {author} {\bibfnamefont {K.}~\bibnamefont {He}}, \bibinfo {author}
  {\bibfnamefont {X.}~\bibnamefont {Ma}}, \bibinfo {author} {\bibfnamefont
  {Q.-K.}\ \bibnamefont {Xue}},  \emph {et~al.},\ }\href@noop {} {\bibfield
  {journal} {\bibinfo  {journal} {Physical Review Letters}\ }\textbf {\bibinfo
  {volume} {114}},\ \bibinfo {pages} {216601} (\bibinfo {year}
  {2015})}\BibitemShut {NoStop}%
\bibitem [{\citenamefont {Minkov}\ \emph {et~al.}(2004)\citenamefont {Minkov},
  \citenamefont {Germanenko},\ and\ \citenamefont
  {Gornyi}}]{minkov2004magnetoresistance}%
  \BibitemOpen
  \bibfield  {author} {\bibinfo {author} {\bibfnamefont {G.}~\bibnamefont
  {Minkov}}, \bibinfo {author} {\bibfnamefont {A.}~\bibnamefont {Germanenko}},
  \ and\ \bibinfo {author} {\bibfnamefont {I.}~\bibnamefont {Gornyi}},\
  }\href@noop {} {\bibfield  {journal} {\bibinfo  {journal} {Physical Review
  B}\ }\textbf {\bibinfo {volume} {70}},\ \bibinfo {pages} {245423} (\bibinfo
  {year} {2004})}\BibitemShut {NoStop}%
\bibitem [{\citenamefont {Kim}\ \emph {et~al.}(2013)\citenamefont {Kim},
  \citenamefont {Syers}, \citenamefont {Butch}, \citenamefont {Paglione},\ and\
  \citenamefont {Fuhrer}}]{kim2013coherent}%
  \BibitemOpen
  \bibfield  {author} {\bibinfo {author} {\bibfnamefont {D.}~\bibnamefont
  {Kim}}, \bibinfo {author} {\bibfnamefont {P.}~\bibnamefont {Syers}}, \bibinfo
  {author} {\bibfnamefont {N.~P.}\ \bibnamefont {Butch}}, \bibinfo {author}
  {\bibfnamefont {J.}~\bibnamefont {Paglione}}, \ and\ \bibinfo {author}
  {\bibfnamefont {M.~S.}\ \bibnamefont {Fuhrer}},\ }\href@noop {} {\bibfield
  {journal} {\bibinfo  {journal} {Nature Communications}\ }\textbf {\bibinfo
  {volume} {4}},\ \bibinfo {pages} {2040} (\bibinfo {year} {2013})}\BibitemShut
  {NoStop}%
\bibitem [{\citenamefont {Lu}\ \emph {et~al.}(2011)\citenamefont {Lu},
  \citenamefont {Shi},\ and\ \citenamefont {Shen}}]{lu2011competition}%
  \BibitemOpen
  \bibfield  {author} {\bibinfo {author} {\bibfnamefont {H.-Z.}\ \bibnamefont
  {Lu}}, \bibinfo {author} {\bibfnamefont {J.}~\bibnamefont {Shi}}, \ and\
  \bibinfo {author} {\bibfnamefont {S.-Q.}\ \bibnamefont {Shen}},\ }\href@noop
  {} {\bibfield  {journal} {\bibinfo  {journal} {Physical Review Letters}\
  }\textbf {\bibinfo {volume} {107}},\ \bibinfo {pages} {076801} (\bibinfo
  {year} {2011})}\BibitemShut {NoStop}%
\bibitem [{\citenamefont {Altshuler}\ \emph {et~al.}(1982)\citenamefont
  {Altshuler}, \citenamefont {Aronov},\ and\ \citenamefont
  {Khmelnitsky}}]{altshuler1982effects}%
  \BibitemOpen
  \bibfield  {author} {\bibinfo {author} {\bibfnamefont {B.~L.}\ \bibnamefont
  {Altshuler}}, \bibinfo {author} {\bibfnamefont {A.}~\bibnamefont {Aronov}}, \
  and\ \bibinfo {author} {\bibfnamefont {D.}~\bibnamefont {Khmelnitsky}},\
  }\href@noop {} {\bibfield  {journal} {\bibinfo  {journal} {Journal of Physics
  C: Solid State Physics}\ }\textbf {\bibinfo {volume} {15}},\ \bibinfo {pages}
  {7367} (\bibinfo {year} {1982})}\BibitemShut {NoStop}%
\bibitem [{\citenamefont {Liao}\ \emph {et~al.}(2017)\citenamefont {Liao},
  \citenamefont {Ou}, \citenamefont {Liu}, \citenamefont {He}, \citenamefont
  {Ma}, \citenamefont {Xue},\ and\ \citenamefont {Li}}]{liao2017enhanced}%
  \BibitemOpen
  \bibfield  {author} {\bibinfo {author} {\bibfnamefont {J.}~\bibnamefont
  {Liao}}, \bibinfo {author} {\bibfnamefont {Y.}~\bibnamefont {Ou}}, \bibinfo
  {author} {\bibfnamefont {H.}~\bibnamefont {Liu}}, \bibinfo {author}
  {\bibfnamefont {K.}~\bibnamefont {He}}, \bibinfo {author} {\bibfnamefont
  {X.}~\bibnamefont {Ma}}, \bibinfo {author} {\bibfnamefont {Q.-K.}\
  \bibnamefont {Xue}}, \ and\ \bibinfo {author} {\bibfnamefont
  {Y.}~\bibnamefont {Li}},\ }\href@noop {} {\bibfield  {journal} {\bibinfo
  {journal} {Nature Communications}\ }\textbf {\bibinfo {volume} {8}},\
  \bibinfo {pages} {16071} (\bibinfo {year} {2017})}\BibitemShut {NoStop}%
\bibitem [{\citenamefont {Br{\"u}ne}\ \emph {et~al.}(2011)\citenamefont
  {Br{\"u}ne}, \citenamefont {Liu}, \citenamefont {Novik}, \citenamefont
  {Hankiewicz}, \citenamefont {Buhmann}, \citenamefont {Chen}, \citenamefont
  {Qi}, \citenamefont {Shen}, \citenamefont {Zhang},\ and\ \citenamefont
  {Molenkamp}}]{brune2011quantum}%
  \BibitemOpen
  \bibfield  {author} {\bibinfo {author} {\bibfnamefont {C.}~\bibnamefont
  {Br{\"u}ne}}, \bibinfo {author} {\bibfnamefont {C.}~\bibnamefont {Liu}},
  \bibinfo {author} {\bibfnamefont {E.}~\bibnamefont {Novik}}, \bibinfo
  {author} {\bibfnamefont {E.}~\bibnamefont {Hankiewicz}}, \bibinfo {author}
  {\bibfnamefont {H.}~\bibnamefont {Buhmann}}, \bibinfo {author} {\bibfnamefont
  {Y.}~\bibnamefont {Chen}}, \bibinfo {author} {\bibfnamefont {X.}~\bibnamefont
  {Qi}}, \bibinfo {author} {\bibfnamefont {Z.}~\bibnamefont {Shen}}, \bibinfo
  {author} {\bibfnamefont {S.}~\bibnamefont {Zhang}}, \ and\ \bibinfo {author}
  {\bibfnamefont {L.}~\bibnamefont {Molenkamp}},\ }\href@noop {} {\bibfield
  {journal} {\bibinfo  {journal} {Physical Review Letters}\ }\textbf {\bibinfo
  {volume} {106}},\ \bibinfo {pages} {126803} (\bibinfo {year}
  {2011})}\BibitemShut {NoStop}%
\bibitem [{\citenamefont {Volkov}\ and\ \citenamefont
  {Pankratov}(1985)}]{volkov1985two}%
  \BibitemOpen
  \bibfield  {author} {\bibinfo {author} {\bibfnamefont {B.}~\bibnamefont
  {Volkov}}\ and\ \bibinfo {author} {\bibfnamefont {O.}~\bibnamefont
  {Pankratov}},\ }\href@noop {} {\bibfield  {journal} {\bibinfo  {journal}
  {JETP Lett}\ }\textbf {\bibinfo {volume} {42}},\ \bibinfo {pages} {178}
  (\bibinfo {year} {1985})}\BibitemShut {NoStop}%
\bibitem [{\citenamefont {Tchoumakov}\ \emph {et~al.}(2017)\citenamefont
  {Tchoumakov}, \citenamefont {Jouffrey}, \citenamefont {Inhofer},
  \citenamefont {Bocquillon}, \citenamefont {Pla{\c{c}}ais}, \citenamefont
  {Carpentier},\ and\ \citenamefont {Goerbig}}]{tchoumakov2017volkov}%
  \BibitemOpen
  \bibfield  {author} {\bibinfo {author} {\bibfnamefont {S.}~\bibnamefont
  {Tchoumakov}}, \bibinfo {author} {\bibfnamefont {V.}~\bibnamefont
  {Jouffrey}}, \bibinfo {author} {\bibfnamefont {A.}~\bibnamefont {Inhofer}},
  \bibinfo {author} {\bibfnamefont {E.}~\bibnamefont {Bocquillon}}, \bibinfo
  {author} {\bibfnamefont {B.}~\bibnamefont {Pla{\c{c}}ais}}, \bibinfo {author}
  {\bibfnamefont {D.}~\bibnamefont {Carpentier}}, \ and\ \bibinfo {author}
  {\bibfnamefont {M.}~\bibnamefont {Goerbig}},\ }\href@noop {} {\bibfield
  {journal} {\bibinfo  {journal} {Physical Review B}\ }\textbf {\bibinfo
  {volume} {96}},\ \bibinfo {pages} {201302} (\bibinfo {year}
  {2017})}\BibitemShut {NoStop}%
\bibitem [{\citenamefont {Inhofer}\ \emph {et~al.}(2017)\citenamefont
  {Inhofer}, \citenamefont {Tchoumakov}, \citenamefont {Assaf}, \citenamefont
  {Feve}, \citenamefont {Berroir}, \citenamefont {Jouffrey}, \citenamefont
  {Carpentier}, \citenamefont {Goerbig}, \citenamefont {Pla{\c{c}}ais},
  \citenamefont {Bendias} \emph {et~al.}}]{inhofer2017observation}%
  \BibitemOpen
  \bibfield  {author} {\bibinfo {author} {\bibfnamefont {A.}~\bibnamefont
  {Inhofer}}, \bibinfo {author} {\bibfnamefont {S.}~\bibnamefont {Tchoumakov}},
  \bibinfo {author} {\bibfnamefont {B.}~\bibnamefont {Assaf}}, \bibinfo
  {author} {\bibfnamefont {G.}~\bibnamefont {Feve}}, \bibinfo {author}
  {\bibfnamefont {J.-M.}\ \bibnamefont {Berroir}}, \bibinfo {author}
  {\bibfnamefont {V.}~\bibnamefont {Jouffrey}}, \bibinfo {author}
  {\bibfnamefont {D.}~\bibnamefont {Carpentier}}, \bibinfo {author}
  {\bibfnamefont {M.}~\bibnamefont {Goerbig}}, \bibinfo {author} {\bibfnamefont
  {B.}~\bibnamefont {Pla{\c{c}}ais}}, \bibinfo {author} {\bibfnamefont
  {K.}~\bibnamefont {Bendias}},  \emph {et~al.},\ }\href@noop {} {\bibfield
  {journal} {\bibinfo  {journal} {Physical Review B}\ }\textbf {\bibinfo
  {volume} {96}},\ \bibinfo {pages} {195104} (\bibinfo {year}
  {2017})}\BibitemShut {NoStop}%
\bibitem [{\citenamefont {Cano}\ \emph {et~al.}(2017)\citenamefont {Cano},
  \citenamefont {Bradlyn}, \citenamefont {Wang}, \citenamefont {Hirschberger},
  \citenamefont {Ong},\ and\ \citenamefont {Bernevig}}]{cano2017chiral}%
  \BibitemOpen
  \bibfield  {author} {\bibinfo {author} {\bibfnamefont {J.}~\bibnamefont
  {Cano}}, \bibinfo {author} {\bibfnamefont {B.}~\bibnamefont {Bradlyn}},
  \bibinfo {author} {\bibfnamefont {Z.}~\bibnamefont {Wang}}, \bibinfo {author}
  {\bibfnamefont {M.}~\bibnamefont {Hirschberger}}, \bibinfo {author}
  {\bibfnamefont {N.~P.}\ \bibnamefont {Ong}}, \ and\ \bibinfo {author}
  {\bibfnamefont {B.~A.}\ \bibnamefont {Bernevig}},\ }\href@noop {} {\bibfield
  {journal} {\bibinfo  {journal} {Physical Review B}\ }\textbf {\bibinfo
  {volume} {95}},\ \bibinfo {pages} {161306} (\bibinfo {year}
  {2017})}\BibitemShut {NoStop}%
\bibitem [{\citenamefont {Chatterjee}\ \emph {et~al.}(2021)\citenamefont
  {Chatterjee}, \citenamefont {Khalid}, \citenamefont {Inbar}, \citenamefont
  {Goswami}, \citenamefont {Guo}, \citenamefont {Chang}, \citenamefont {Young},
  \citenamefont {Fedorov}, \citenamefont {Read}, \citenamefont {Janotti} \emph
  {et~al.}}]{chatterjee2021controlling}%
  \BibitemOpen
  \bibfield  {author} {\bibinfo {author} {\bibfnamefont {S.}~\bibnamefont
  {Chatterjee}}, \bibinfo {author} {\bibfnamefont {S.}~\bibnamefont {Khalid}},
  \bibinfo {author} {\bibfnamefont {H.~S.}\ \bibnamefont {Inbar}}, \bibinfo
  {author} {\bibfnamefont {A.}~\bibnamefont {Goswami}}, \bibinfo {author}
  {\bibfnamefont {T.}~\bibnamefont {Guo}}, \bibinfo {author} {\bibfnamefont
  {Y.-H.}\ \bibnamefont {Chang}}, \bibinfo {author} {\bibfnamefont
  {E.}~\bibnamefont {Young}}, \bibinfo {author} {\bibfnamefont {A.~V.}\
  \bibnamefont {Fedorov}}, \bibinfo {author} {\bibfnamefont {D.}~\bibnamefont
  {Read}}, \bibinfo {author} {\bibfnamefont {A.}~\bibnamefont {Janotti}},
  \emph {et~al.},\ }\href@noop {} {\bibfield  {journal} {\bibinfo  {journal}
  {Science Advances}\ }\textbf {\bibinfo {volume} {7}},\ \bibinfo {pages}
  {eabe8971} (\bibinfo {year} {2021})}\BibitemShut {NoStop}%
\bibitem [{\citenamefont {Skinner}\ \emph {et~al.}(2012)\citenamefont
  {Skinner}, \citenamefont {Chen},\ and\ \citenamefont
  {Shklovskii}}]{skinner2012bulk}%
  \BibitemOpen
  \bibfield  {author} {\bibinfo {author} {\bibfnamefont {B.}~\bibnamefont
  {Skinner}}, \bibinfo {author} {\bibfnamefont {T.}~\bibnamefont {Chen}}, \
  and\ \bibinfo {author} {\bibfnamefont {B.}~\bibnamefont {Shklovskii}},\
  }\href@noop {} {\bibfield  {journal} {\bibinfo  {journal} {Physical review
  letters}\ }\textbf {\bibinfo {volume} {109}},\ \bibinfo {pages} {176801}
  (\bibinfo {year} {2012})}\BibitemShut {NoStop}%
\bibitem [{\citenamefont {Kresse}\ and\ \citenamefont
  {Furthmüller}(1996)}]{VASP}%
  \BibitemOpen
  \bibfield  {author} {\bibinfo {author} {\bibfnamefont {G.}~\bibnamefont
  {Kresse}}\ and\ \bibinfo {author} {\bibfnamefont {J.}~\bibnamefont
  {Furthmüller}},\ }\href {\doibase
  https://doi.org/10.1016/0927-0256(96)00008-0} {\bibfield  {journal} {\bibinfo
   {journal} {Computational Materials Science}\ }\textbf {\bibinfo {volume}
  {6}},\ \bibinfo {pages} {15 } (\bibinfo {year} {1996})}\BibitemShut {NoStop}%
\bibitem [{\citenamefont {Perdew}\ \emph {et~al.}(1996)\citenamefont {Perdew},
  \citenamefont {Burke},\ and\ \citenamefont {Ernzerhof}}]{PBE}%
  \BibitemOpen
  \bibfield  {author} {\bibinfo {author} {\bibfnamefont {J.~P.}\ \bibnamefont
  {Perdew}}, \bibinfo {author} {\bibfnamefont {K.}~\bibnamefont {Burke}}, \
  and\ \bibinfo {author} {\bibfnamefont {M.}~\bibnamefont {Ernzerhof}},\ }\href
  {\doibase 10.1103/PhysRevLett.77.3865} {\bibfield  {journal} {\bibinfo
  {journal} {Phys. Rev. Lett.}\ }\textbf {\bibinfo {volume} {77}},\ \bibinfo
  {pages} {3865} (\bibinfo {year} {1996})}\BibitemShut {NoStop}%
\bibitem [{\citenamefont {Monkhorst}\ and\ \citenamefont
  {Pack}(1976)}]{PhysRevB.13.5188}%
  \BibitemOpen
  \bibfield  {author} {\bibinfo {author} {\bibfnamefont {H.~J.}\ \bibnamefont
  {Monkhorst}}\ and\ \bibinfo {author} {\bibfnamefont {J.~D.}\ \bibnamefont
  {Pack}},\ }\href {\doibase 10.1103/PhysRevB.13.5188} {\bibfield  {journal}
  {\bibinfo  {journal} {Phys. Rev. B}\ }\textbf {\bibinfo {volume} {13}},\
  \bibinfo {pages} {5188} (\bibinfo {year} {1976})}\BibitemShut {NoStop}%
\bibitem [{\citenamefont {Bl\"ochl}(1994)}]{PAW}%
  \BibitemOpen
  \bibfield  {author} {\bibinfo {author} {\bibfnamefont {P.~E.}\ \bibnamefont
  {Bl\"ochl}},\ }\href {\doibase 10.1103/PhysRevB.50.17953} {\bibfield
  {journal} {\bibinfo  {journal} {Phys. Rev. B}\ }\textbf {\bibinfo {volume}
  {50}},\ \bibinfo {pages} {17953} (\bibinfo {year} {1994})}\BibitemShut
  {NoStop}%
\bibitem [{\citenamefont {Seibel}\ \emph {et~al.}(2015)\citenamefont {Seibel},
  \citenamefont {Schoop}, \citenamefont {Xie}, \citenamefont {Gibson},
  \citenamefont {Webb}, \citenamefont {Fuccillo}, \citenamefont {Krizan},\ and\
  \citenamefont {Cava}}]{seibel2015gold}%
  \BibitemOpen
  \bibfield  {author} {\bibinfo {author} {\bibfnamefont {E.~M.}\ \bibnamefont
  {Seibel}}, \bibinfo {author} {\bibfnamefont {L.~M.}\ \bibnamefont {Schoop}},
  \bibinfo {author} {\bibfnamefont {W.}~\bibnamefont {Xie}}, \bibinfo {author}
  {\bibfnamefont {Q.~D.}\ \bibnamefont {Gibson}}, \bibinfo {author}
  {\bibfnamefont {J.~B.}\ \bibnamefont {Webb}}, \bibinfo {author}
  {\bibfnamefont {M.~K.}\ \bibnamefont {Fuccillo}}, \bibinfo {author}
  {\bibfnamefont {J.~W.}\ \bibnamefont {Krizan}}, \ and\ \bibinfo {author}
  {\bibfnamefont {R.~J.}\ \bibnamefont {Cava}},\ }\href@noop {} {\bibfield
  {journal} {\bibinfo  {journal} {Journal of the American Chemical Society}\
  }\textbf {\bibinfo {volume} {137}},\ \bibinfo {pages} {1282} (\bibinfo {year}
  {2015})}\BibitemShut {NoStop}%
\bibitem [{\citenamefont {Strohbeen}\ \emph {et~al.}(2019)\citenamefont
  {Strohbeen}, \citenamefont {Du}, \citenamefont {Zhang}, \citenamefont
  {Shourov}, \citenamefont {Rodolakis}, \citenamefont {McChesney},
  \citenamefont {Voyles},\ and\ \citenamefont
  {Kawasaki}}]{strohbeen2019electronically}%
  \BibitemOpen
  \bibfield  {author} {\bibinfo {author} {\bibfnamefont {P.~J.}\ \bibnamefont
  {Strohbeen}}, \bibinfo {author} {\bibfnamefont {D.}~\bibnamefont {Du}},
  \bibinfo {author} {\bibfnamefont {C.}~\bibnamefont {Zhang}}, \bibinfo
  {author} {\bibfnamefont {E.~H.}\ \bibnamefont {Shourov}}, \bibinfo {author}
  {\bibfnamefont {F.}~\bibnamefont {Rodolakis}}, \bibinfo {author}
  {\bibfnamefont {J.~L.}\ \bibnamefont {McChesney}}, \bibinfo {author}
  {\bibfnamefont {P.~M.}\ \bibnamefont {Voyles}}, \ and\ \bibinfo {author}
  {\bibfnamefont {J.~K.}\ \bibnamefont {Kawasaki}},\ }\href@noop {} {\bibfield
  {journal} {\bibinfo  {journal} {Physical Review Materials}\ }\textbf
  {\bibinfo {volume} {3}},\ \bibinfo {pages} {024201} (\bibinfo {year}
  {2019})}\BibitemShut {NoStop}%
\bibitem [{\citenamefont {Dresselhaus}(1955)}]{dresselhaus1955spin}%
  \BibitemOpen
  \bibfield  {author} {\bibinfo {author} {\bibfnamefont {G.}~\bibnamefont
  {Dresselhaus}},\ }\href@noop {} {\bibfield  {journal} {\bibinfo  {journal}
  {Physical Review}\ }\textbf {\bibinfo {volume} {100}},\ \bibinfo {pages}
  {580} (\bibinfo {year} {1955})}\BibitemShut {NoStop}%
\bibitem [{\citenamefont {Brydon}\ \emph {et~al.}(2016)\citenamefont {Brydon},
  \citenamefont {Wang}, \citenamefont {Weinert},\ and\ \citenamefont
  {Agterberg}}]{brydon2016pairing}%
  \BibitemOpen
  \bibfield  {author} {\bibinfo {author} {\bibfnamefont {P.}~\bibnamefont
  {Brydon}}, \bibinfo {author} {\bibfnamefont {L.}~\bibnamefont {Wang}},
  \bibinfo {author} {\bibfnamefont {M.}~\bibnamefont {Weinert}}, \ and\
  \bibinfo {author} {\bibfnamefont {D.}~\bibnamefont {Agterberg}},\ }\href@noop
  {} {\bibfield  {journal} {\bibinfo  {journal} {Physical Review Letters}\
  }\textbf {\bibinfo {volume} {116}},\ \bibinfo {pages} {177001} (\bibinfo
  {year} {2016})}\BibitemShut {NoStop}%
\bibitem [{\citenamefont {Potter}\ \emph {et~al.}(2014)\citenamefont {Potter},
  \citenamefont {Kimchi},\ and\ \citenamefont
  {Vishwanath}}]{potter2014quantum}%
  \BibitemOpen
  \bibfield  {author} {\bibinfo {author} {\bibfnamefont {A.~C.}\ \bibnamefont
  {Potter}}, \bibinfo {author} {\bibfnamefont {I.}~\bibnamefont {Kimchi}}, \
  and\ \bibinfo {author} {\bibfnamefont {A.}~\bibnamefont {Vishwanath}},\
  }\href@noop {} {\bibfield  {journal} {\bibinfo  {journal} {Nature
  Communications}\ }\textbf {\bibinfo {volume} {5}},\ \bibinfo {pages} {5161}
  (\bibinfo {year} {2014})}\BibitemShut {NoStop}%
\bibitem [{\citenamefont {Davis}\ \emph {et~al.}(1999)\citenamefont {Davis},
  \citenamefont {Jones}, \citenamefont {Falkenberg}, \citenamefont {Seehofer},
  \citenamefont {Johnson},\ and\ \citenamefont
  {McConville}}]{davis1999evidence}%
  \BibitemOpen
  \bibfield  {author} {\bibinfo {author} {\bibfnamefont {A.}~\bibnamefont
  {Davis}}, \bibinfo {author} {\bibfnamefont {R.}~\bibnamefont {Jones}},
  \bibinfo {author} {\bibfnamefont {G.}~\bibnamefont {Falkenberg}}, \bibinfo
  {author} {\bibfnamefont {L.}~\bibnamefont {Seehofer}}, \bibinfo {author}
  {\bibfnamefont {R.}~\bibnamefont {Johnson}}, \ and\ \bibinfo {author}
  {\bibfnamefont {C.~F.}\ \bibnamefont {McConville}},\ }\href@noop {}
  {\bibfield  {journal} {\bibinfo  {journal} {Applied Physics Letters}\
  }\textbf {\bibinfo {volume} {75}},\ \bibinfo {pages} {1938} (\bibinfo {year}
  {1999})}\BibitemShut {NoStop}%
\bibitem [{\citenamefont {Liu}\ and\ \citenamefont
  {Santos}(1994)}]{liu1994surface}%
  \BibitemOpen
  \bibfield  {author} {\bibinfo {author} {\bibfnamefont {W.}~\bibnamefont
  {Liu}}\ and\ \bibinfo {author} {\bibfnamefont {M.}~\bibnamefont {Santos}},\
  }\href@noop {} {\bibfield  {journal} {\bibinfo  {journal} {Surface Science}\
  }\textbf {\bibinfo {volume} {319}},\ \bibinfo {pages} {172} (\bibinfo {year}
  {1994})}\BibitemShut {NoStop}%
\bibitem [{nex()}]{nextnano}%
  \BibitemOpen
  \href@noop {} {}\bibinfo {note}
  {\lowercase{h}ttps://www.nextnano.de/}\BibitemShut {NoStop}%
\end{thebibliography}

\end{document}